\documentclass[prx,aps, 11pt, onecolumn, superscriptaddress, notitlepage,nofootinbib, floatfix, longbibliography]{revtex4-1}
\usepackage[utf8]{inputenc}
\usepackage[usenames, dvipsnames]{xcolor}
\usepackage{dsfont}
\usepackage{amsmath}
\usepackage{amsfonts}
\usepackage{amssymb}
\usepackage{graphicx}
\graphicspath{{./figs/}}
\usepackage[pdfencoding=auto, psdextra]{hyperref} 
\hypersetup{linktocpage}
\hypersetup{colorlinks=true,citecolor=quantumpurple,linkcolor=quantumpurple, urlcolor=quantumpurple}
\usepackage{natbib}
\usepackage{environ}
\usepackage{enumitem}
\usepackage{mathrsfs}
\usepackage{changepage}
\usepackage[caption=false]{subfig}
\usepackage{makecell}

\NewEnviron{eqs}{%
\begin{equation}\begin{split}
    \BODY
\end{split}\end{equation}
}

\makeatletter 
\def\l@subsubsection#1#2{}
\makeatother

\hypersetup{colorlinks=true,citecolor=blue,linkcolor=blue, urlcolor=blue}

\newcommand{\ZZ}{\mathbb{Z}}
\newcommand{\ra}{\rangle}
\newcommand{\la}{\langle}

\usepackage{physics}
\usepackage{faktor}

\newcommand{\pop}{\hat{p}}
\newcommand{\xop}{\hat{x}}
\newcommand{\cA}{\mathcal{A}}
\newcommand{\cC}{\mathcal{C}}

\newcommand{\cB}{\mathcal{B}}
\newcommand{\cM}{\mathcal{M}}
\newcommand{\cS}{\mathcal{S}}

\newcommand{\cL}{\mathcal{L}}
\newcommand{\cH}{\mathcal{H}}
\newcommand{\cW}{\mathcal{W}}

\newcommand{\bZ}{\mathbb{Z}}
\newcommand{\bR}{\mathbb{R}}
\newcommand{\bC}{\mathbb{C}}
\newcommand{\bL}{\mathbb{L}}

\def\U{\mathrm{U}(1)}
\newcommand{\CS}[2]{$\U_{#1} \times \U_{#2}$}

\DeclareMathOperator{\spn}{span}

\newtheorem{definition}{Definition}

\definecolor{quantumpurple}{RGB}{82, 35, 124}
\definecolor{teal}{RGB}{0,128,128}

\begin{document}

\title{{Topological stabilizer models on continuous variables}}

\author{Julio C. Magdalena de la Fuente}
\email{jm@juliomagdalena.de}
\affiliation{Dahlem Center for Complex Quantum Systems, Freie Universit\"at Berlin, 14195 Berlin, Germany}
\author{{Tyler D.~Ellison}}
\email{tdelliso@purdue.edu}
\affiliation{Perimeter Institute for Theoretical Physics, Waterloo, Ontario N2L 2Y5, Canada}
\affiliation{Department of Physics, Yale University, New Haven, Connecticut  06511-8499, USA}
\affiliation{Department of Physics and Astronomy, Purdue University, West Lafayette, IN, 47907}
\affiliation{Purdue Quantum Science and Engineering Institute, Purdue University, West Lafayette, IN, 47907}
\author{Meng Cheng}
\affiliation{Department of Physics, Yale University, New Haven, Connecticut  06511-8499, USA}
\author{Dominic J. Williamson}
\affiliation{School of Physics, University of Sydney, Sydney, New South Wales 2006, Australia}

\begin{abstract}
\noindent 
\footnotesize
We construct a family of two-dimensional topological stabilizer codes on continuous variable (CV) degrees of freedom, which generalize homological rotor codes and the toric-GKP code. Our topological codes are built using the concept of boson condensation -- we start from a parent stabilizer code based on an $\mathbb{R}$ gauge theory and condense various bosonic excitations. This produces a large class of topological CV stabilizer codes, including ones that are characterized by the anyon theories of \CS{2n}{-2m} Chern-Simons theories, for arbitrary pairs of positive integers $(n,m)$.
Most notably, this includes anyon theories that are non-chiral and nevertheless do not admit a gapped boundary.
It is widely believed that such anyon theories cannot be realized by any stabilizer model on finite-dimensional systems.
We conjecture that these CV codes go beyond codes obtained from concatenating a topological qudit code with a local encoding into CVs, and thus, constitute the first example of topological codes that are intrinsic to CV systems.
Moreover, we study the Hamiltonians associated to the topological 
CV stabilizer codes and show that, although they have a gapless spectrum, they can become gapped with the addition of a quadratic perturbation.
We show that similar methods can be used to construct a gapped Hamiltonian whose anyon theory agrees with a $\U_2$ Chern-Simons theory.
Our work initiates the study of scalable stabilizer codes that are intrinsic to CV systems and highlights how error-correcting codes can be used to design and analyze many-body systems of CVs that model lattice gauge theories.

\end{abstract}

\maketitle

\tableofcontents

\section{Introduction}

Quantum error correction is essential for storing and processing quantum information so that it is protected from faulty operations and decoherence induced by the environment~\cite{Preskill2018quantumcomputing, shor1997faulttolerantquantumcomputation, Dennis2002quantummemory}. Many of the quantum error-correcting (QEC) schemes developed thus far employ qubits as their fundamental building blocks, which may arise from, for example, hyperfine states in neutral atoms or trapped ions~\cite{Bluvstein2024logical,Quantinuum2023racetrack}, or the lowest two energy levels of a superconducting circuit~\cite{Google2023scaling,Google2024quantumerrorcorrectionsurface}. These systems, however, generically support a much larger Hilbert space than the two levels that are actively used for quantum error correction.~A promising approach to minimize the physical overhead of quantum computation is thus to tailor QEC codes to hardware to take full advantage of the available state space.

This motivates the use of continuous variable (CV) degrees of freedom for processing quantum information, which for example, arise from a superconducting qubit in a microwave cavity~\cite{Cai2021bosonic, Noh2021bosonic, Puri2021bosonic}, the motional states of a trapped ion~\cite{fossfeig2024trappedionsimulation, Gutierrez2017motional, Gan2020Hybrid}, or the rotational degrees of freedom of a molecule~\cite{Albert2020molecule}.
The Hilbert space of a CV is equivalent to the state space of a harmonic oscillator and as such, is formally infinite dimensional. CVs thus promise to be a hardware efficient alternative to processing quantum information with qubits.

The majority of QEC codes developed for CV systems have been designed for a single CV, referred to as a single mode~\cite{GKP2001, Albert2014cat, Albert2016binomial, Albert2018singlemode,Grimmsmo2020rotation}. There is by now a rich family of single-mode QEC codes, and they have been used to demonstrate the protection of quantum information beyond break-even~\cite{Sivak2023breakeven}. Despite this progress, single-mode QEC codes ultimately offer only limited protection of the encoded quantum information.
They are not able to achieve the logical error rates needed to perform complex quantum computations, such as factoring into large prime numbers~\cite{Gidney2021howtofactorbit}.  

To suppress the logical error rate further, we need to develop \textit{scalable} QEC codes, i.e., codes with a prescription for introducing additional degrees of freedom so as to obtain arbitrarily low logical error rates. There has been development of multi-mode CV codes, some of which are scalable~\cite{Harrington2001Achievable, Royer2022Encoding, Jain2024, Denys_2023, Conrad2022, Raynal_2010}. However, there are limited explicit examples of scalable codes, and they lack desirable properties for physical implementation, such as geometrically local checks.

Topological QEC codes, for which the code space is defined by the ground state subspace of a local Hamiltonian, are a paradigmatic class of scalable codes with geometrically local checks~\cite{Kitaev2003fault-tolerant}. Topological QEC codes also offer robust protection from geometrically local errors and can be implemented in planar geometries~\cite{Dennis2002quantummemory, bravyi1998surfacecode}. They furthermore admit a well-understood set of natively fault-tolerant logical gates, and sophisticated decoders which yield high error thresholds~\cite{Bombin2010Twist,Litinski2019gameofsurfacecodes,Fowler2012surface}. 

To develop scalable CV codes with local checks, it is thus natural to consider topological QEC codes built from CVs. One approach is to leverage known topological QEC codes for systems of qudits. Specifically, one can begin by encoding a qudit into each CV using a single-CV Gottesman-Kitaev-Preskill (GKP) code~\cite{GKP2001}. Then, one can implement any known topological QEC code on the encoded qudits. In other words, a single-CV GKP code can be concatenated with a topological code. In the case of concatenating a GKP code with the toric code, this gives the toric-GKP code, which was first benchmarked in Ref.~\cite{Vuillot2019toricGKP}. Although this approach is promising and indeed yields scalable QEC codes, it is unsatisfying in that each CV is used in isolation to encode a qudit and the scalability is coming entirely from the scalability of the qudit code.

Developing topological CV QEC codes beyond concatenation with qudit codes has proven to be a subtle problem.
Refs.~\cite{Niset2009nogo,Vuillot2019toricGKP,Hanggli_2022}
established no-go theorems which state that CVs cannot be encoded fault-tolerantly into systems of physical CVs.\footnote{Throughout this work, fault-tolerant encodings refer to those that have an error-correcting threshold in the presence of local errors and faulty measurements.} 
Ref.~\cite{vuillot2023homological} then explored building topological QEC codes out of rotors, which can be thought of as a CV with a large energetic constraint.
The rotor codes introduced in Ref.~\cite{vuillot2023homological} have the benefit that they fundamentally rely on the infinite-dimensional state space of each rotor.
However, despite encoding a discrete number of qubits on homologically nontrivial manifolds, they unfortunately fail to be fault tolerant without additional concatenation, as they do not admit an increasing code distance for a reasonable set of local errors. 

In this work, we generalize the toric-GKP code and homological rotor codes of Refs.~\cite{Vuillot2019toricGKP} and \cite{vuillot2023homological} to construct fault-tolerant topological CV codes, which we expect are beyond code concatenation -- in the sense that they are not simply a local encoding of a qudit into CVs concatenated with a stabilizer code on qudits. This expectation is based on the particular Abelian anyon theories exhibited by the topological stabilizer codes. Specifically, the anyon theories include those captured by \CS{2n}{-2m} Chern-Simons (CS) theories~\cite{wen2004quantum, Cano2014bulkedge, Yuan-Ming2012CSinteger, Ellison2022stabilizer}. This is noteworthy since, for example, the theory parameterized by $(n,m)=(1,2)$ is Witt nontrivial -- that is, the anyon theory does not admit a gapped boundary, despite being non-chiral~\cite{Davydov2013a,LevinEdge,Barkeshli2013Classification}. The prevailing expectation is that such theories do not admit a description in terms of a qudit stabilizer model, thus putting the topological CV stabilizer code beyond code concatenation. 

We construct our new examples of topological CV stabilizer codes starting with a CV stabilizer code based on an $\bR$ gauge theory. This parent stabilizer code encodes two CVs, so in accordance with the no-go theorems of Refs.~\cite{Niset2009nogo,Vuillot2019toricGKP,Hanggli_2022}, it is not fault tolerant. However, from this model, we build fault-tolerant topological stabilizer codes by implementing boson condensation~\cite{Burnell2018anyon}. Boson condensation has proven to be a valuable tool for both building new QEC codes~\cite{Ellison2022stabilizer, Ellison2022subsystem, Dua2024rewinding} and understanding the structures of familiar topological codes~\cite{Kesselring2024condensation}. This work can thus be seen as generalizing constructions of QEC codes based on boson condensation to systems with infinite-dimensional degrees of freedom.

The parent stabilizer code has bosonic excitations that can be interpreted as the gauge charges, gauge fluxes, and certain charge-flux composites in an $\bR$ gauge theory. When the bosonic excitations are proliferated -- by for example, measuring the short string operators that create the excitations -- this generically yields new topological QEC codes.
For example, condensing bare charges and fluxes in the $\bR$ gauge theory produces the rotor codes of Ref.~\cite{vuillot2023homological} or the toric-GKP code of Ref.~\cite{Vuillot2019toricGKP}. 

On the other hand, condensing bosons formed from charge-flux composites gives rise to a much more diverse set of topological codes, including those characterized by \CS{2n}{-2m} CS theories. More generally, we show that this approach is capable of producing topological codes characterized by non-chiral $\U\times\U$ CS theories for $2 \times 2$ $K$ matrices with all even entries. 
Condensing charge-flux composites moreover gives a new mechanism for encoding discrete quantum information in CVs and sidesteps the issue faced by the codes in Ref.~\cite{vuillot2023homological}, in which a logical string operator that moves a pure flux can be smeared across the full system by applying stabilizers -- thereby decreasing the distance when the code size is increased. 

The organization of the paper is as follows. In Section~\ref{sec:example2-4}, we get started by providing an example of a topological CV stabilizer code, called the $K_{2,-4}$ stabilizer code. This stabilizer code is characterized by a \CS{2}{-4} CS theory, which corresponds to an anyon theory that does not admit a gapped boundary and is beyond known qudit stabilizer codes. To prove that the code is topological, we exploit a particular factorization of the Hilbert space according to the $\U_2$ and $\U_{-4}$ factors of the CS theory, generalizing the invertible subalgebras of Ref.~\cite{Haah2023invertible}. We then further show that, although the associated Hamiltonian is gapless, a local quadratic perturbation is sufficient to open up a gap for a system without boundary -- thus, giving a gapped model of the \CS{2}{-4} CS theory with a $8$-fold ground state degeneracy on a torus. We additionally construct a controllably-solvable, gapped Hamiltonian for the anyon theory of a $\U_2$ CS theory. To the best of our knowledge, this is the first example of such a model defined on a tensor-product Hilbert space with short-range, physically realistic interactions.

In Section~\ref{sec:condensation}, we first formally extend the existing defintions of topological stabilizer codes to CVs, and then describe our general construction of topological CV codes, starting from an $\bR$ gauge theory.
In Section~\ref{sec:Examples}, we describe examples of topological CV stabilizer codes constructed by either (i) condensing a single type of boson, yielding the homological rotor codes and examples characterized by $\U_{2n}$ CS theories, or (ii) consecutively condensing two types of bosons, which yields codes associated to $\ZZ_N$ toric codes and non-chiral CS theories with a $2\times 2$ $K$ matrix with even entries. 
We highlight the example where two consecutive condensations lead to models that admit the same anyon theory as \CS{2n}{-2m} CS theories. 
In Section~\ref{sec:Discussion}, we discuss our results and potential future directions.
In Appendix~\ref{app:twisted_hoppingterms}, we elaborate on the algebra of operators that appears in the condensation of flux-charge composites. 
In Appendix~\ref{app:twisted-condensation-group}, we show how the single condensation leads to a finite subtheory of gapped excitations characterized by $\U_{2n}$ CS theory.

\noindent \textit{Note -- } A work~\cite{Victor2024tiger}, which appeared shortly after this work, studies many-body generalizations of various cat and binomial codes, which, similar to the expectation for the codes presented in this work, are not obtained by concatenation. 

\section{Example: $K_{2,-4}$ stabilizer model}\label{sec:example2-4}

We begin with an example that illustrates the topological stabilizer models of Section~\ref{sec:condensation} and \ref{sec:Examples}. We refer to the example in this section as the $K_{2,-4}$ stabilizer model, where the notation $K_{2,-4}$ represents the fact that the gapped excitations of the model capture the anyon theory of a \CS{2}{-4} CS theory.\footnote{Therefore the $K$ matrix is $K=\begin{pmatrix}
    2 & 0 \\
    0 & -4
\end{pmatrix}$.} We leave a complete description of the anyon theory for a later section, but for now, note that this anyon theory is not captured by any known stabilizer model or commuting projector Hamiltonian on qudits. One unique property of this model is that, on a torus, it has an 8-fold ground state degeneracy. This is in contrast to all known topological stabilizer models on qudits, where the dimension of the code space on a torus is a square~\cite{Ellison2022stabilizer}.

We start by defining the Hilbert space and the $K_{2,-4}$ stabilizer Hamiltonian in Section~\ref{sec: K24 hilbert space and code}. We then discuss the excitations of the Hamiltonian and emphasize that there are gapped point-like excitations that exhibit the properties of a \CS{2}{-4} CS anyon theory. Accordingly, the model has an 8-fold ground state degeneracy on a torus, which we verify in Section~\ref{sec: K24 splitting of algebra}. In Section~\ref{sec: K24 splitting of algebra}, we also prove that the associated stabilizer code is topological by employing a particular non-spatial factorization of the Hilbert space. Finally, in Section~\ref{sec: K24 gapping bulk} we demonstrate that, although the $K_{2,-4}$ Hamiltonian admits gapless excitations, local perturbations are sufficient to open up a gap -- yielding a gapped (non-commuting) Hamiltonian for the \CS{2}{-4} CS anyon theory. We similarly construct a gapped model for a $\U_2$ CS theory.

\subsection{Hilbert space and Hamiltonian} \label{sec: K24 hilbert space and code}

The $K_{2,-4}$ stabilizer Hamiltonian is defined on a square lattice with a CV on every edge. This is to say that, for every edge $e$, there is an associated position operator $\xop_e$ and momentum operator $\pop_e$, which satisfy the canonical commutation relations:
\begin{align}
    \comm{\xop_e}{\pop_{e'}} = 
    \begin{cases}
        i & e = e', \\
        0 & e \neq e'.
    \end{cases}
\end{align}
We further define the displacement operators $X_e$ and $Z_e$ at the edge $e$ as:
\begin{align}\label{eq:displacements-def}
    X_e = e^{-i \pop_e}, \quad Z_e = e^{i \xop_e}.
\end{align}
These satisfy the following commutation relations, for $s,t \in \bR$:
\begin{align} \label{eq: commutation relations}
    Z_e^tX_{e'}^s = 
    \begin{cases}
        e^{ist} X_{e'}^sZ_e^t & e=e', \\
        \,\,\,\,\,\,\,\,\, X_{e'}^sZ_e^t & e\neq e'.
    \end{cases}
\end{align}
The displacement operators are natural generalizations of the Pauli operators on qubits, so as such, we occasionally refer to products of displacement operators as Pauli operators. We note that, in contrast to Pauli operators on qudits, the displacement operators have infinite order. 

We are now prepared to introduce the $K_{2,-4}$ stabilizer Hamiltonian, denoted by $H_{2,-4}$. We first note that the Hamiltonian decomposes into a sum of two Hamiltonians $H_2$ and $H_{-4}$:
\begin{align} \label{eq: K24 Hamiltonian}
    H_{2,-4} = H_2 + H_{-4}.
\end{align}
$H_2$ and $H_{-4}$ are further composed of vertex terms and edge terms. Explicitly, the Hamiltonians are:
\begin{eqs} \label{eq: K24 H2H4 decomposition}
    H_2 &= -\sum_v A_v^{(2)} - \sum_e C_e^{(2)} + \text{h.c.}, \\
    H_{-4} &= -\sum_v A_v^{(-4)} - \sum_e C_e^{(-4)} + \text{h.c.}
\end{eqs}
The vertex terms $A_v^{(2)}$ and edge terms $C_e^{(2)}$ for $H_2$ are graphically represented as
\begin{align} \label{eq: 2 stabilizers}
    A_v^{(2)} = \vcenter{\hbox{\includegraphics[scale=0.3]{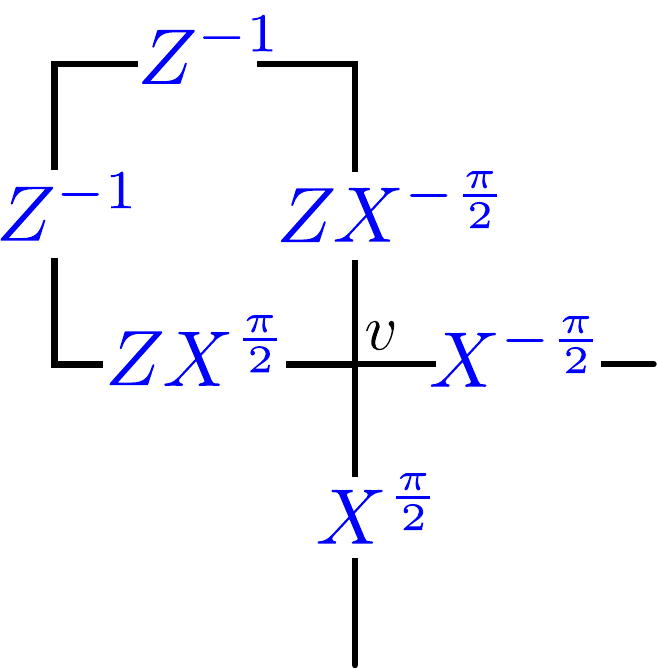}}}, \quad C_e^{(2)} = \vcenter{\hbox{\includegraphics[scale=0.3]{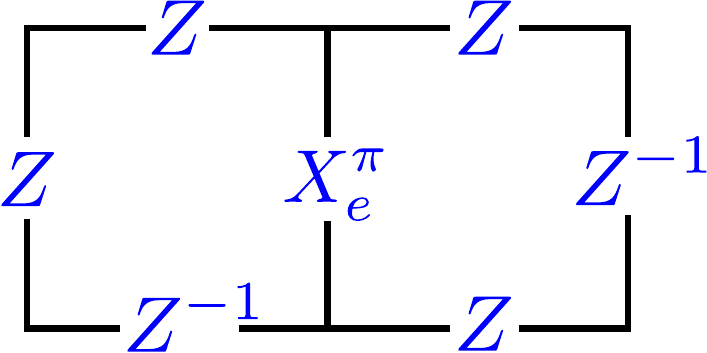}}}\,,\,\,\, \vcenter{\hbox{\includegraphics[scale=0.3]{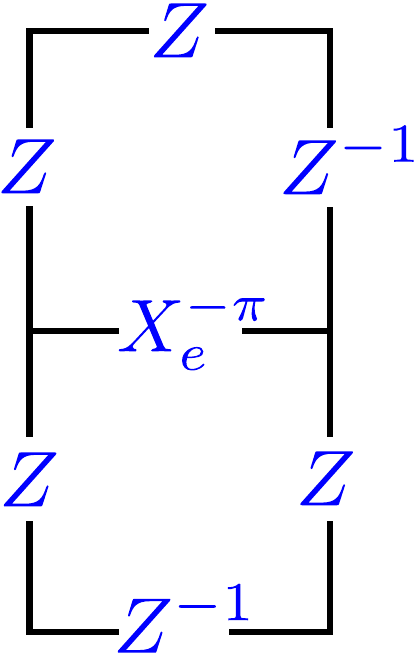}}}.
\end{align}
Here and throughout, the figures should be implicitly understood as products of single-site Pauli operators. Likewise, the vertex terms $A_v^{(-4)}$ and edge terms $C_e^{(-4)}$ of $H_{-4}$ are:
\begin{align} \label{eq: -4 stabilizers}
    A_v^{(-4)} = \vcenter{\hbox{\includegraphics[scale=0.3]{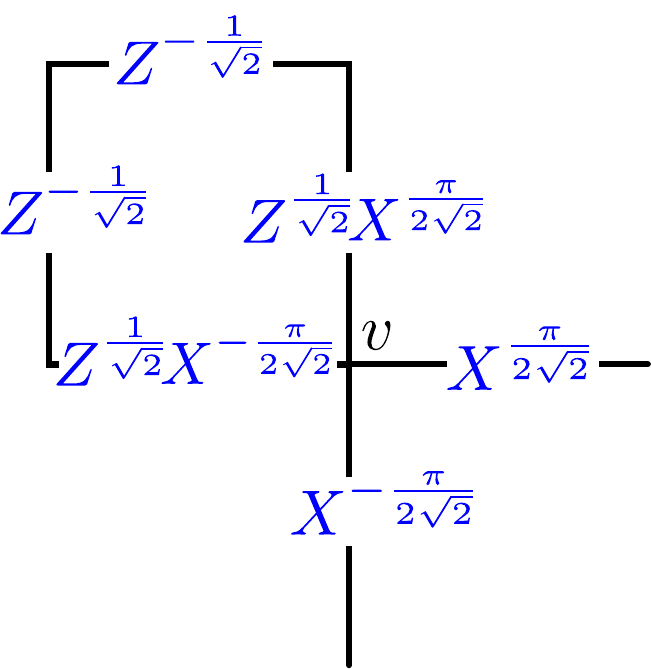}}}, \quad C_e^{(-4)} = \vcenter{\hbox{\includegraphics[scale=0.3]{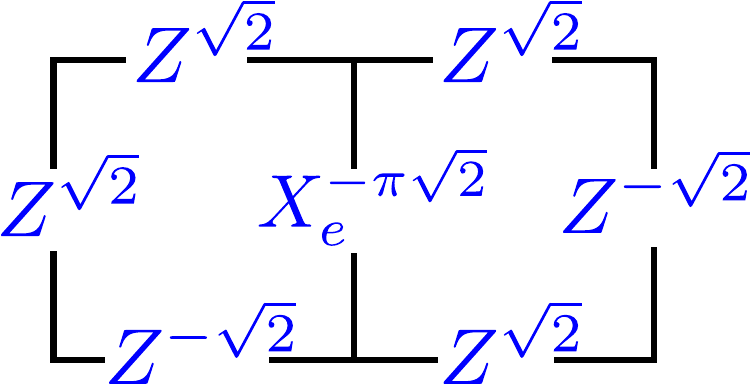}}}\,,\,\,\, \vcenter{\hbox{\includegraphics[scale=0.3]{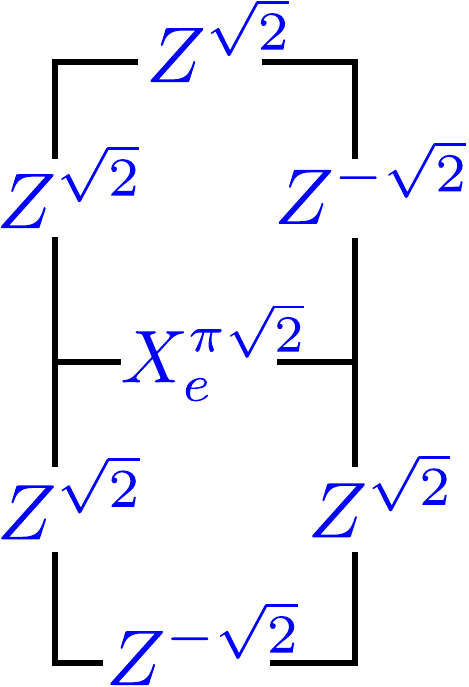}}}.
\end{align}

It can be checked, using Eq.~\eqref{eq: commutation relations}, that the Hamiltonian terms are all mutually commuting. Therefore, the ground state subspace is spanned by the mutual $+1$ eigenstates of the Hamiltonian terms. This allows us to define a stabilizer group $\cS_{2,-4}$ generated by the Hamiltonian terms of $H_{2,-4}$. We refer to the resulting stabilizer code as the $K_{2,-4}$ stabilizer code. It is convenient to define the following two subgroups, which generate $\cS_{2,-4}$:
\begin{eqs}
    \cS_2 &= \la \{A_{v}^{(2)}, \, C_{e}^{(2)} \, | \, v \in V, \, e \in E\}  \ra, \\ 
    \cS_{-4} &= \la \{A_{v}^{(-4)}, \, C_{e}^{(-4)} \, | \, v \in V, \, e \in E \} \ra,
\end{eqs}
where $V$ and $E$ are the sets of vertices and edges, respectively. The angled bracket notation here denotes the fact that the stabilizer subgroups are generated by integer powers of the arguments. 

\begin{figure*}[t]
\centering
{\includegraphics[width=1\textwidth]{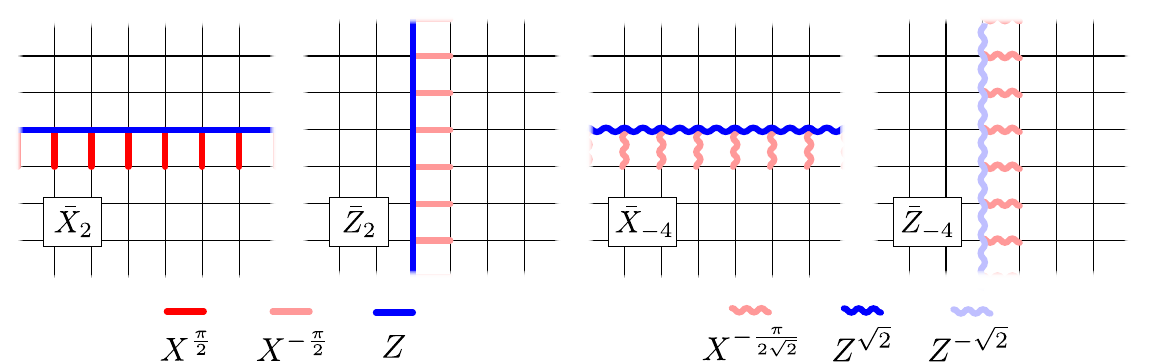}}
\caption{The logical operators of the $K_{2,-4}$ stabilizer code on a torus. The logical operators $\bar{X}_2$ and $\bar{Z}_2$ square to a product of stabilizers in $\cS_2$ and anticommute. Therefore, they define an encoding of a qubit. Likewise the fourth powers of $\bar{X}_{-4}$ and $\bar{Z}_{-4}$ are products of stabilizers in $\cS_{-4}$ and they fail to commute by a fourth root of unity. Hence, they define an encoding of a four-dimensional qudit.  }
\label{fig: K24 logicals}
\end{figure*}

By construction, the ground state subspace of the Hamiltonian coincides with the code space of the $K_{2,-4}$ stabilizer code. On a torus, the $K_{2,-4}$ stabilizer code encodes one qubit and one four-dimensional qudit. In Fig.~\ref{fig: K24 logicals}, we show a representation of the logical operators. In general, the logical operators are string operators that wrap around the torus and can be deformed to topologically equivalent paths by multiplication with  stabilizers. We prove in the next section that the code space is indeed 8-dimensional and show that the $K_{2,-4}$ stabilizer code is a topological CV stabilizer code. This is to say that the stabilizer group is locally generated and there are no local representations of the logical operators.\\

\noindent \textbf{Remark:} The ground states of the Hamiltonian $H_{2,-4}$ are not normalizable -- that is, they do not belong to the space of square-integrable states $L^2(\bR)^{\otimes N}$, assuming that there are $N$ CVs. To make this explicit, we first note that the stabilizer group $\cS_{2,-4}$ can be understood as a multi-mode GKP code, as defined in Ref.~\cite{GKP2001}. Any such multi-mode GKP code is then unitarily equivalent to a collection of independent single-mode GKP codes~\cite[Cor. 1]{Conrad2022}. The ground states of a single-mode GKP code are sums over delta functions in the $\xop$ basis, peaked at integer multiples of specific values of $x$. Therefore, the ground states of a single-mode GKP code, and hence the ground states of $H_{2,-4}$, are not normalizable.\footnote{Moreover, unlike plane waves, the code states of an $N$-mode GKP code are not simple distributions over $L^2(\bR)^{\otimes N}$. This is despite the fact that they are sums of delta functions, each of which is well-defined distribution on $L^2(\bR)$. For a single-mode GKP code, this can be seen by evaluating the code state on a wave function with an amplitude that goes as $1/\abs{x}$ for large $x$. This does not evaluate to a finite value, due to the divergence of the sum $\sum_{k}1/k$. One way to give precise mathematical meaning to GKP code states is as {tempered distributions}. These are distributions on the space of Schwartz functions, i.e., smooth functions that decay faster than any polynomial for large $\abs{x}$.} \\

First, however, we explore the excitations of the $K_{2,-4}$ stabilizer Hamiltonian. The excitations of the $K_{2,-4}$ stabilizer Hamiltonian can be generated using the short string operators $W_e^{(2)}$ and $W_e^{(-4)}$. The short string operators $W_e^{(2)}$ and $W_e^{(-4)}$ for an edge $e$ are graphically represented as: 
\begin{subequations}
\begin{eqs} \label{eq: K24 short strings 2}
    W_e^{(2)} = \vcenter{\hbox{\includegraphics[scale=0.3]{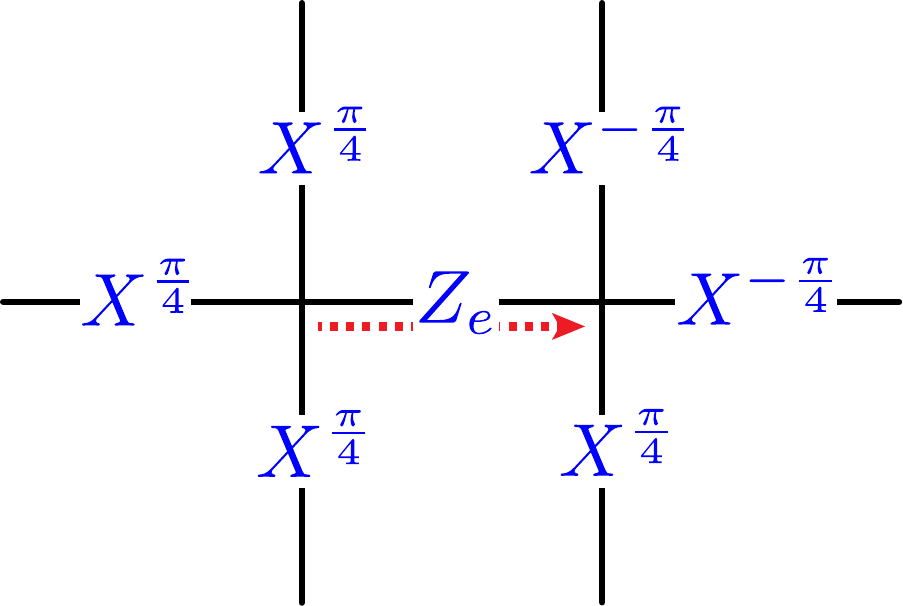}}}\,, \quad \vcenter{\hbox{\includegraphics[scale=0.3]{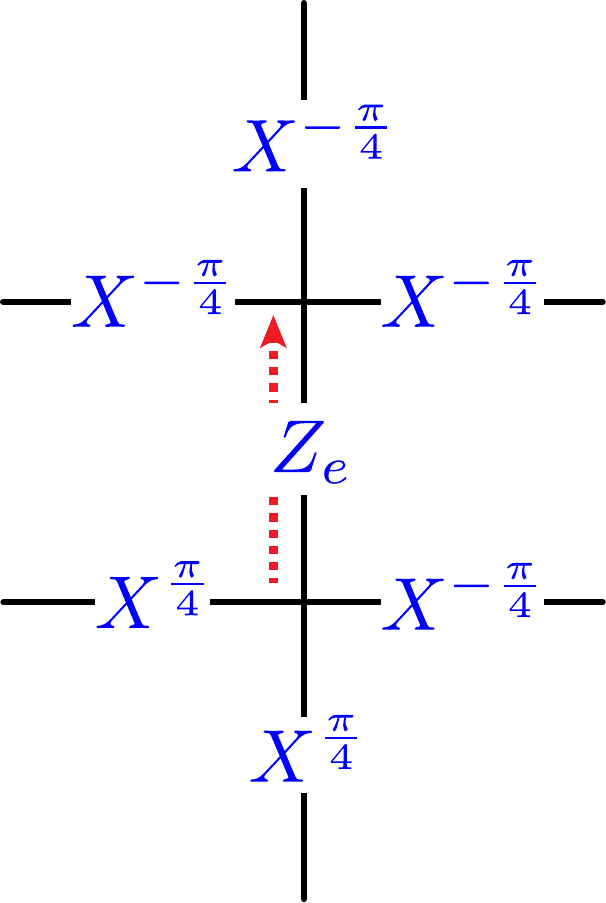}}}\,,
\end{eqs}
\begin{eqs} \label{eq: K24 short strings 4}
    W_e^{(-4)} = \vcenter{\hbox{\includegraphics[scale=0.3]{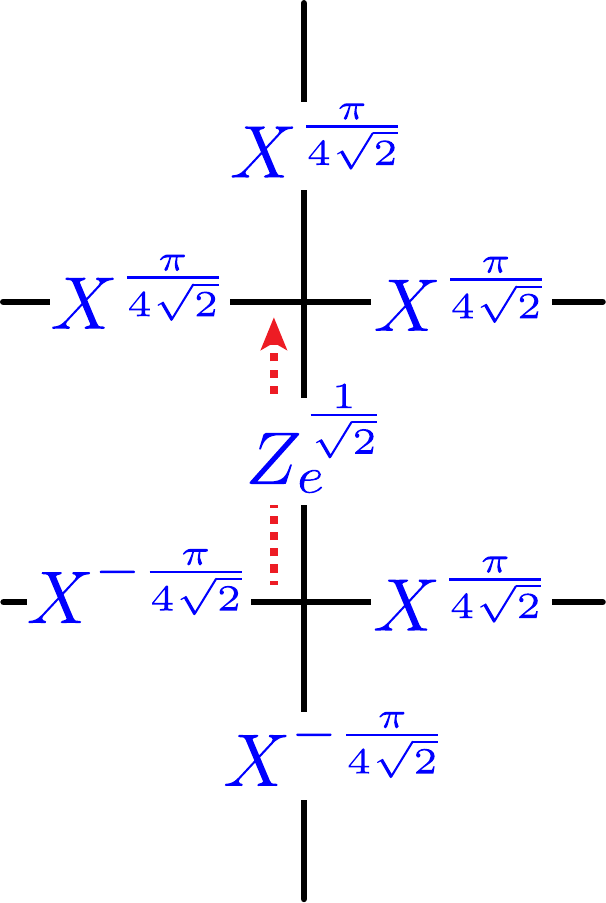}}}\,, \quad \vcenter{\hbox{\includegraphics[scale=0.3]{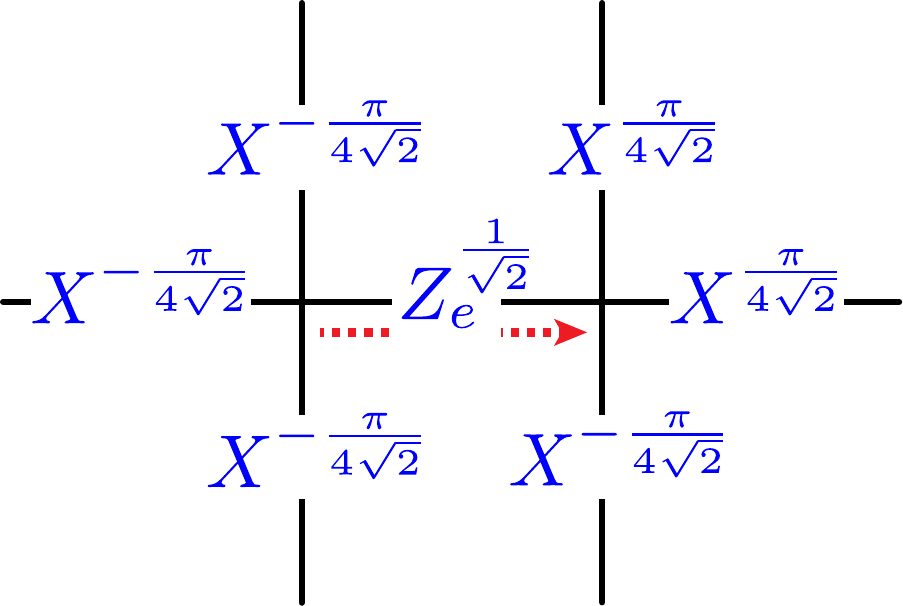}}},
\end{eqs}    
\end{subequations}
where the red dashed arrows denote an orientation of the short string operator. The opposite orientation is obtained by Hermitian conjugation.

To characterize the excitations, we further define long string operators from products of the short string operators. For any oriented path $\gamma$, we define the string operators:
\begin{align} \label{eq: K24 string operators}
    W_{\gamma}^{(2)} = \prod_{e \in \gamma} \vec{W}_e^{(2)}, \quad W_{\gamma}^{(-4)} = \prod_{e \in \gamma} \vec{W}_e^{(-4)}.
\end{align}
Here, the notation $\vec{W}_e^{(2)}$ and $\vec{W}_e^{(-4)}$ should be understood as the Hermitian conjugation of the short string operators in Eqs.~\eqref{eq: K24 short strings 2} and \eqref{eq: K24 short strings 4} only if the orientation is anti-aligned with the orientation of $\gamma$. Note that the short string operators are not all mutually commuting, so the ordering of the products in Eqs.~\eqref{eq: K24 short strings 2} and \eqref{eq: K24 short strings 4} is not generally well defined.\footnote{Note that the string operators are only ambiguous up to a phase, which does not effect the excitations that they create.} However, one can take as a convention that the short string operators are ordered according to the orientation of $\gamma$. 

\begin{figure*}[t]
\centering
{\includegraphics[width=.9\textwidth]{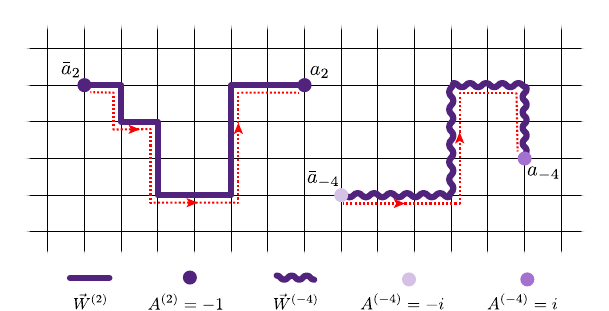}}
\caption{Anyonic excitations of the $K_{2,-4}$ stabilizer Hamiltonian. The $K_{2,-4}$ stabilizer Hamiltonian admits point-like excitations that cannot be created or destroyed with operators localized near the excitations. The excitations are instead created at the endpoints of string operators (purple) along oriented paths (red dashed). Only the vertex terms $A_v^{(2)}$ and $A_v^{(-4)}$ are violated at the endpoints of the strings.}
\label{fig: K24 excitations}
\end{figure*}

We now explore the excitations created by the string operators in Eq.~\eqref{eq: K24 string operators} along an open path $\gamma$, as shown in Fig.~\ref{fig: K24 excitations}. It can be checked that the string operators $W_\gamma^{(2)}$ and $W_\gamma^{(-4)}$ commute with all of the Hamiltonian terms along the length of $\gamma$ and only fail to commute with isolated vertex terms at the endpoints. Moreover, the commutation relations of the string operator with the vertex terms at the endpoints cannot be reproduced using operators that are only supported in the vicinity of the endpoints. In this sense, the local excitations cannot be created or destroyed using operators localized to the endpoints and thus belong to nontrivial superselection sectors -- analogous to anyonic excitations. 

Formally, we can treat these gapped excitations as anyons and compute their characteristic properties, i.e., their fusion rules, exchange statistics, and braiding relations. We refer to the anyons created at the endpoints of $\gamma$ by $W_{\gamma}^{(2)}$ as $\bar{a}_2$ and $a_2$, where the orientation of the path $\gamma$ points from $\bar{a}_2$ to $a_2$. Here, $\bar{a}_2$ denotes the inverse of $a_2$, i.e., composing two string operators of the same type from head to tail leaves no excitations along the length of the long string. Likewise, we refer to the anyons created by $W_{\gamma}^{(-4)}$ as $\bar{a}_{-4}$ and $a_{-4}$.

The fusion rules of the anyons can be determined by taking integer powers of the string operators. It can be checked that $W_e^{(2)}$ squares to a product of two $A_v^{(2)}$ stabilizers and a $C_e^{(2)}$ stabilizer, as represented graphically in Eq.~\eqref{eq: K24 2 fusion}.
Therefore, the short string operator $\left(W_e^{(2)}\right)^2$ does not create any excitations.
Similarly, the fourth power of $W_e^{(-4)}$ is a product of stabilizers.
These relations imply the fusion rules:
\begin{align}
    a_2^2 = 1, \quad a_{-4}^4 = 1. 
\end{align}
Thus, the anyons $a_2$ and $a_{-4}$ generate a $\ZZ_2 \times \ZZ_4$ group under fusion. 

Next, the exchange statistics and braiding relations can be determined using the standard T-junction calculation~\cite{Levin2003Fermions,Haah2018classification}.\footnote{Since the anyon theory is Abelian, the braiding relations $B(a,b)$ of $a$ and $b$ can be determined from the exchange statistics by $B(a,b) = \theta(ab)/\theta(a)\theta(b)$.} We find that $a_2$ and $a_{-4}$ braid trivially with each other: 
\begin{align}
    B(a_2, a_{-4}) = 1.
\end{align}
Furthermore, their exchange statistics are:
\begin{align}
    \theta(a_2) = i, \quad \theta(a_{-4}) = -i\sqrt{i}.
\end{align}
In other words, exchanging two $a_2$ or $a_{-4}$ anyons produces a phase of $i$ or $-i\sqrt{i}$, respectively. 

For Abelian anyon theories, the anyon theory is fully determined by the fusion rules, the exchange statistics, and braiding relations of the generators. Therefore, we can conclude that the anyon theory generated by $a_2$ and $a_{-4}$ is precisely the anyon theory of a \CS{2}{-4} CS theory. It follows from the arguments in Section~\ref{sec: K24 splitting of algebra} that $a_2$ and $a_{-4}$ exhaust all of the anyons in the system. 

The stabilizers and logical operators of the $K_{2,-4}$ stabilizer code can be understood in terms of the anyons $a_2$ and $a_{-4}$. The vertex stabilizers $A_v^{(2)}$ and $A_v^{(-4)}$ are small, counterclockwise loops of string operator for $a_2$ and $a_{-4}$, respectively. Moreover, the logical operators in Fig.~\ref{fig: K24 logicals} are string operators of $a_2$ and $a_{-4}$, wrapped around non-contractible paths of the torus.
Hence, the logical operators are created by moving the anyons around the torus, and their commutation relations are the result of the braiding of the anyons. \\

\noindent \textbf{Remark:} Note that the anyon theory of a \CS{2}{-4} CS theory does not contain any nontrivial bosons, and hence does not admit a gappable boundary~\cite{LevinEdge, Saulina2011boundary}.
There is strong evidence and it is widely believed that any anyon theory of a (gapped) commuting projector Hamiltonian on DVs must have a gappable boundary~\cite{Kitaev_2006,Haah2023QCA,Haah2018classification,ruba2025wittgroups}.
This suggests that there is no local unitary that maps $\cS_{2,-4}$ to the stabilizer group of a concatenated code.
Eqs.~\eqref{eq: 2 stabilizers} and \eqref{eq: -4 stabilizers} provide further evidence for this conjecture based on number-theoretic properties of the displacement vectors defining the stabilizer group. In particular, the stabilizer group
$\cS_{-4}$ is generated by displacements by multiples of $\sqrt{2}$, while $\cS_{2}$ does not contain displacements of that form.
The $A_v^{(-4)}$ terms commute because the irrational phases that come from individual CVs cancel and the $A_v^{(2)}$ terms because rational phases cancel.
Since this commutation structure cannot be directly reproduced on any DV system, it seems implausible that there is a local unitary mapping $\cS_{2,-4}$ to a stabilizer group obtained from concatenating a DV code with a local CV code.\\

Besides the gapped excitations described above, the Hamiltonian $H_{2,-4}$ also admits gapless excitations.
To create the gapless excitations, we consider the string operators $\left(W_\gamma^{(2)}\right)^s$ and $\left( W_\gamma^{(-4)} \right)^s$, for $s$ valued in $\bR$. One can check that, if $s \notin \ZZ$, the string operators fail to commute with the vertex terms at their endpoints and also fail to commute with the edge terms along the length of $\gamma$. The string operators $\left(W_\gamma^{(2)}\right)^s$ and $\left( W_\gamma^{(-4)} \right)^s$ are able to create excitations with arbitrarily small energies as $s$ is tuned infinitesimally above zero.

To gain physical intuition for the excitations created by the string operators with $s \notin \ZZ$, let us consider the string operators on an open path $\gamma$. For $s \notin \ZZ$, the string operators create point-like excitations at their endpoints in the sense that they violate the vertex terms. However, since the string operators fail to commute with the edge terms along their length, the point-like excitations are linearly confined. This is in contrast to the anyonic excitations $a_2$ and $a_{-4}$, which are deconfined. As a consequence of the confinement, the excitations with continuously parameterized energies do not play a significant role in the stabilizer code. 

\subsection{Factorization of the Pauli group} \label{sec: K24 splitting of algebra}

To argue that the $K_{2,-4}$ stabilizer code is topological and moreover that the logical operators in Fig.~\ref{fig: K24 logicals} exhaust all of the logical operators on a torus, we exploit a particular factorization of the group of Pauli operators. The factorization follows from the short string operators in Eqs.~\eqref{eq: K24 short strings 2} and \eqref{eq: K24 short strings 4}. To make this factorization explicit, we define $\cW_2$ and $\cW_{-4}$ as the subgroups of Pauli operators generated by $\bR$-valued powers of the $W_e^{(2)}$ and $W_e^{(-4)}$ short string operators, respectively:
\begin{eqs}
    \cW_2 &= \left \langle \left( W_e^{(2)} \right)^s \,|\, e \in E, \, s \in \bR \right \rangle  \\
    \cW_{-4} &= \left \langle \left( W_e^{(-4)} \right)^s \,|\, e \in E, \, s \in \bR \right \rangle.
\end{eqs}

We argue now that any Pauli operator can be expressed as $W_2 \times W_{-4}$, where $W_2 \in \cW_2$ and $W_{-4}\in \cW_{-4}$. It is insightful to first note that the stabilizer group of the $K_{2,-4}$ stabilizer code splits according to this factorization. That is, the subgroups $\cS_2$ and $\cS_{-4}$ are contained within $\cW_2$ and $\cW_{-4}$, respectively. To see this explicitly, we note that the vertex stabilizers $A_v^{(2)}$ and $A_v^{(-4)}$ are small counterclockwise loops of string operator:
\begin{align}
    A_v^{(2)} \propto \vcenter{\hbox{\includegraphics[scale=0.3]{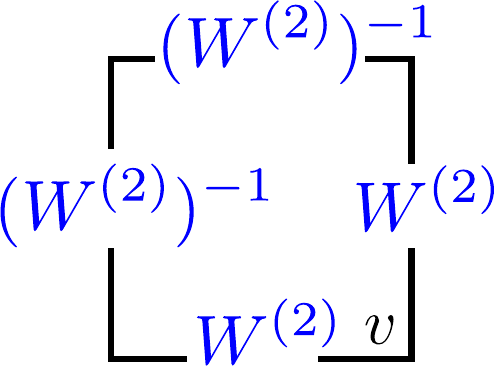}}}\,, \quad 
    A_v^{(-4)} \propto \vcenter{\hbox{\includegraphics[scale=0.3]{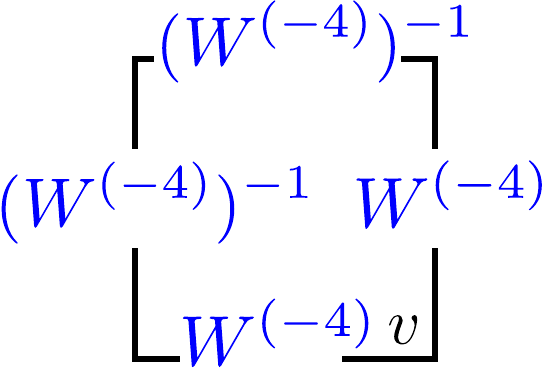}}}\,.
\end{align}
Furthermore, the edge stabilizers are given by:
\begin{eqs} \label{eq: K24 Ce in terms of We}
    C_e^{(2)} &\propto \vcenter{\hbox{\includegraphics[scale=0.3]{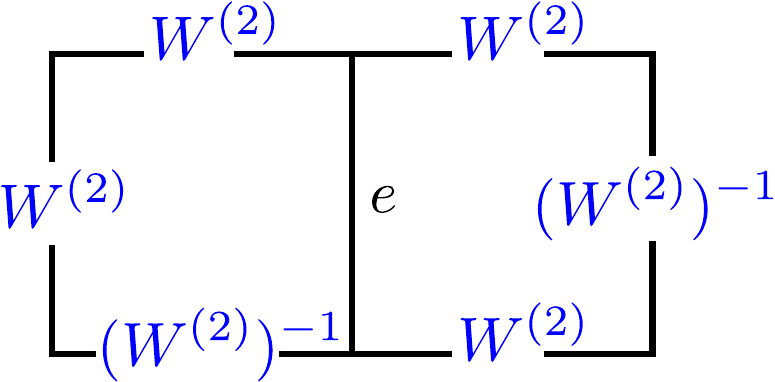}}}\,, \,\,\,\,\,\,\,\,\,\,\,\vcenter{\hbox{\includegraphics[scale=0.3]{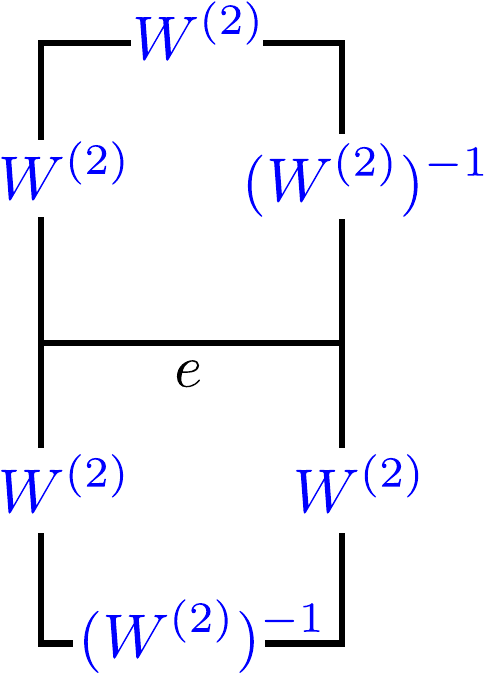}}}, \\
    C_e^{(-4)} &\propto \vcenter{\hbox{\includegraphics[scale=0.3]{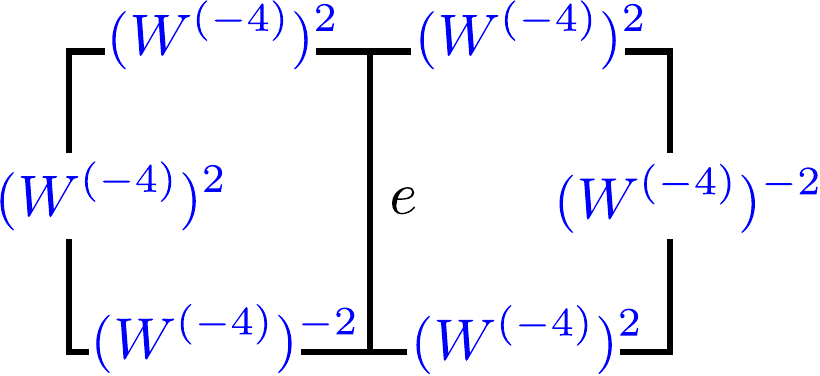}}}\,, \,\,\,\, \vcenter{\hbox{\includegraphics[scale=0.3]{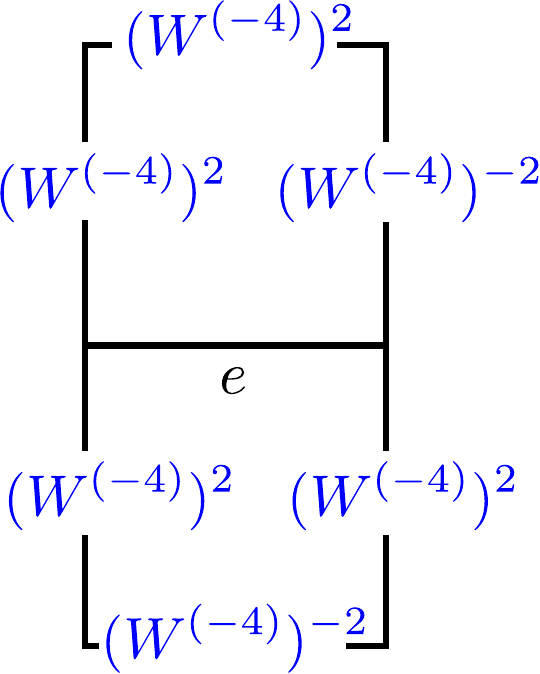}}}.
\end{eqs}

More generally, the factorization follows from two properties of the short string operators. First, the group of Pauli operators is contained within $\cW_2 \times \cW_{-4}$. In particular, $X$ and $Z$ are given by:
\begin{align}
    X_e^{o_e}=\left( C_e^{(2)} \right)^{\frac{1}{2\pi}}\left( C_e^{(-4)} \right)^{-\frac{1}{2\sqrt{2}\pi}}, \quad Z_e=\left( W_e^{(2)} \right)^{\frac{1}{2}}\left( W_e^{(-4)} \right)^{\frac{1}{\sqrt{2}}},
\end{align}
where $o_e$ is $+1$ ($-1$) if $e$ is vertical (horizontal), and $C_e^{(2)}$ and $C_e^{(-4)}$ are expressed in terms of $W_e^{(2)}$ and $W_e^{(-4)}$ in Eq.~\eqref{eq: K24 Ce in terms of We}.
Second, for any choice of $s, t \in \bR$ and for every pair of edges $e$ and $e'$, the short string operators satisfy the commutation relations:
\begin{align}
    \left(W_e^{(2)}\right)^s \left(W_{e'}^{(-4)}\right)^t = \left(W_{e'}^{(-4)}\right)^t \left(W_e^{(2)}\right)^s.
\end{align}
Therefore, the groups $\cW_2$ and $\cW_{-4}$ are independent, analogous to Pauli operators on two disjoint subsystems. In fact, it can further be shown that the sums of $\xop$ and $\pop$ operators that generate $W_e^{(2)}$ and $W_e^{(-4)}$ define a factorization of the Hilbert space into non-spatial subsystems. 

One consequence of the splitting of the Pauli group is that the elements of $\cW_2$ ($\cW_{-4}$) necessarily commute with all of the stabilizers in $\cS_{-4}$ ($\cS_2$).
Therefore, the properties of the stabilizer code can be studied by considering $\cS_2$ and $\cS_{-4}$ as independent stabilizer codes over the Pauli groups $\cW_2$ and $\cW_{-4}$, respectively. We use this fact to argue below that the $K_{2,-4}$ stabilizer code is topological. In particular, we show that the stabilizer subgroup $\cS_2$ defines an encoding of a qubit and is topological with respect to the group $\cW_2$. Analogous arguments show that the $\cS_{-4}$ subgroup encodes a 4-dimensional qudit and is topological with respect to $\cW_{-4}$.

We focus now on the subgroup $\cS_2$ over the group $\cW_2$ and search for logical operators generated by elements of $\cW_2$. 
In general, the elements of $\cW_2$ fail to commute with the stabilizers in $\cS_2$. For example, if $s \in \bR$, then the operator $\left(W_e^{(2)}\right)^s$ fails to commute with an edge stabilizer except for when $s \in \ZZ$. Thus, the logical operators in $\cW_2$ must be composed of integer powers of $W_e^{(2)}$. If $k \in \ZZ$, however, then the operator $\left(W_e^{(2)}\right)^k$ fails to commute with a pair of vertex stabilizers except for when $k \in 2\ZZ$. Therefore, any operator in $\cW_2$ that commutes with all of the stabilizers in $\cS_2$ must be a product of: (i) operators $\left(W_e^{(2)}\right)^k$, for $k \in 2\ZZ$, and (ii) integer power of $W_e^{(2)}$ along oriented closed paths. 
We notice, however, that the operator $\left(W_e^{(2)}\right)^k$, for $k \in 2\ZZ$, is a stabilizer belonging to $\cS_2$. This can be seen from the relation:
\begin{align} \label{eq: K24 2 fusion}
    \left(W_e^{(2)}\right)^2 = \vcenter{\hbox{\includegraphics[scale=0.3]{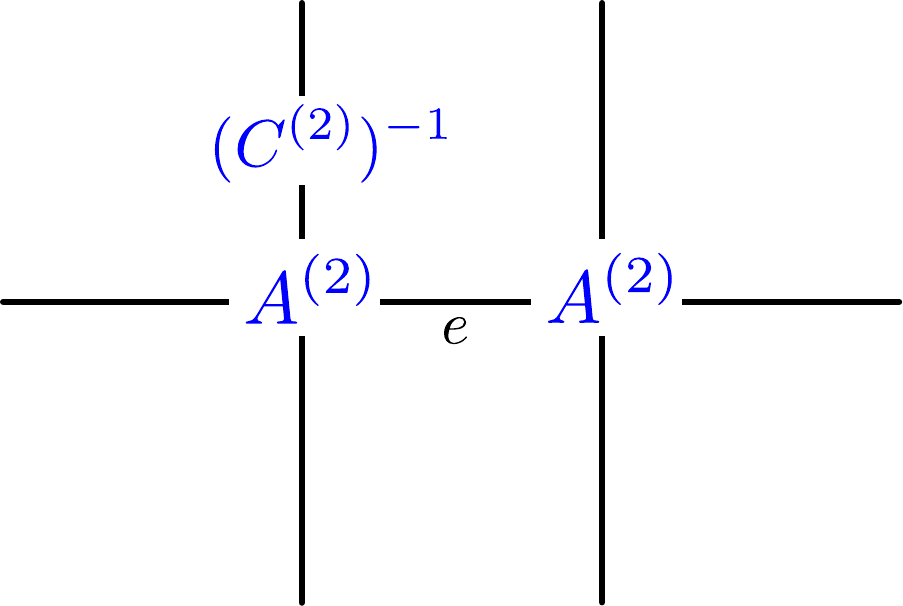}}}\,\,, \,\, \,\,\vcenter{\hbox{\includegraphics[scale=0.3]{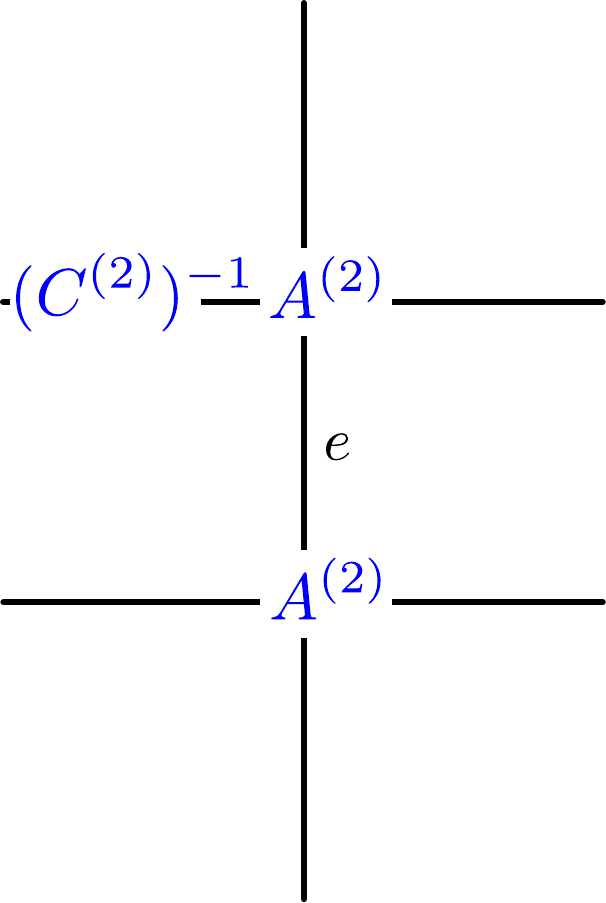}}}.
\end{align}
Likewise, any integer power of $W_e^{(2)}$ along a contractible loop is a product of $A_v^{(2)}$ stabilizers. 

Therefore, the only Pauli operators in $\cW_2$ that commute with all of the stabilizers and are not themselves stabilizers are the non-contractible loops of $W_e^{(2)}$. These are precisely the two leftmost logical operators in Fig.~\ref{fig: K24 logicals}. This further implies that there are no local logical operators generated by elements of $\cW_2$. Similar statements hold for the subgroup $\cS_{-4}$, so the $K_{2,-4}$ stabilizer code is indeed a topological CV stabilizer code that encodes one qubit and one four-dimensional qudit. 

\subsection{Gapping out the bulk} \label{sec: K24 gapping bulk}

In stark contrast to qudit stabilizer models, the Hamiltonian $H_{2,-4}$ is gapless. The spectrum can be gapped out, however, by adding quadratic perturbations. In this section, we argue, more specifically, that adding a small quadratic term to the Hamiltonian $H_{2,-4}$ opens a gap and realizes the $\U_2\times \U_{-4}$ topological order. We then show that, if we add quadratic terms to the Hamiltonian $H_2$, defined in Eq.~\eqref{eq: K24 H2H4 decomposition}, we realize a model whose anyon content is that of the $\U_2$ CS theory.\\

\subsubsection{Gapped Hamiltonian for $\U_2\times \U_{-4}$}
\label{sec:gapping 1}
We define the perturbed Hamiltonian as:
\begin{equation}
\begin{split}
    H'_{2,-4}=H_{2,-4}+H_0,
\end{split}
\end{equation}
with the quadratic perturbation $H_0$ defined as
\begin{eqs}
    H_0=\alpha\sum_{e} \Big[(\hat{w}_{e}^{(2)})^2+(\hat{w}_{e}^{(-4)})^2\Big],
\end{eqs}
for $\alpha >0$. Here, the operators $\hat{w}_e^{(k)}$ are the quadrature of $W_e^{(k)}$, i.e., $W^{(k)}_e=e^{i\hat{w}^{(k)}_e}$. Similarly, we define $\hat{c}_e^{(k)}$ as the quadrature of $C^{(k)}_e$. It can be checked straightforwardly that the two operators satisfy
\begin{equation}
\begin{gathered}
    \comm{\hat{c}_e^{(k)}}{\hat{w}_{e'}^{(k')}} = \pm 2\pi i\delta_{ee'}\delta_{kk'},
\end{gathered}
\label{eq:commu-relation-wc}
\end{equation}
where $\pm$ depends on the orientation of $e,e'$ and the signs of $k,k'$. Importantly, this means that $\hat{w}_e^{(k)}$ commutes with the Hamiltonian $H_{k'}$ for $k \neq k'$.

To see that the perturbation is sufficient to gap out the Hamiltonian, it is convenient to first split the Hamiltonian into vertex terms and edge terms:
\begin{align}
    H'_{2,-4}=H_A+H_C+H_0,
\end{align}
where $H_A$ and $H_C$ are:
\begin{eqs}
    H_A&=-U'\sum_{e}(A_{e}^{(2)}+A_{e}^{(-4)}), \\
    H_C&=-U\sum_{e}(C_{e}^{(2)}+C_{e}^{(-4)}).
\end{eqs}
Following the method developed in Ref.~\cite{Ganeshan2016largecosine}, we first analyze the Hamiltonian $H_C+H_0$ and show that it has a fully gapped spectrum. The effective ground state subspace is extensively degenerate, however, without the vertex terms. We then treat $H_A$ as a perturbation, which further lifts the extensive degeneracy leaving only a constant number of ground states.

To make this explicit, we review the formalism of Ref.~\cite{Ganeshan2016largecosine}. For a CV with $\{\xop_i, \pop_i\}_{i=1}^N$, the formalism in Ref.~\cite{Ganeshan2016largecosine} perturbatively solves the Hamiltonian of the following form
\begin{equation}
    H=-U\sum_{i=1}^M \cos S_i+H_0
\end{equation}
in the limit $U\rightarrow \infty$.
Here, $H_0$ is quadratic and the operators $S_i$ are linear in the $\xop_i$ and $\pop_i$ operators. We further need to assume that (i) the $S_i$ operators are linearly independent, and (ii) the commutator $\frac{1}{2\pi i}[S_i, S_j]$ is an integer, so $[\cos S_i, \cos S_j]=0$. 

The Hamiltonian $H$ gives rise to an effective low-energy subspace $\mathcal{H}_\mathrm{eff}$ defined by ${\cos S_i=1}$, for all $i$.
This low-energy space is separated from other excited states by an energy gap $\Delta$. To specify $\Delta$, we define an $M\times M$ matrix
\begin{equation}
    \mathcal{N}_{ij}=-\frac{1}{4\pi^2}[S_i,[S_j, H_0]].
\end{equation}
It can be checked that $\mathcal{N}$ is a real symmetric matrix and is semi-positive definite. To carry out the perturbative analysis, we require that $\mathcal{N}$ is positive-definite. In that case, we denote the minimal eigenvalue of $\mathcal{N}$ by $\lambda_{\rm min}$. According to Ref.~\cite{Ganeshan2016largecosine}, the gap is then given by $\Delta \sim \sqrt{U\lambda_{\rm min}}$.

For our Hamiltonian, the $c_e^{(k)}$ operators play the role of the $S_i$ operators. Due to the commutation relations in Eq.~\eqref{eq:commu-relation-wc} the $\mathcal{N}$ matrix is diagonal. In fact, the $\U_2$ and $\U_{-4}$ parts completely decouple, so $\mathcal{N}$ is already block-diagonalized. This allows us to focus on one of the blocks, e.g., the block associated to the $\U_2$ layer. We find
\begin{equation}
    \mathcal{N}_{ee'}=\frac{\alpha}{\pi}\delta_{ee'}.
\end{equation}
Hence, the spectrum is independent of the system size, and the gap is $\Delta\sim \sqrt{U\alpha}$.

We can now describe the subspace $\mathcal{H}_\mathrm{eff}$. To this end, define an $M\times M$ matrix $\mathcal{Z}$ by
\begin{equation}
    \mathcal{Z}_{ij}=\frac{1}{2\pi i}[S_i, S_j].
\end{equation}
It is integer, skew-symmetric, and completely determines the structure of the subspace $\mathcal{H}_{\rm eff}$. In particular, the dimension of the space is given by $\sqrt{|\det \mathcal{Z}|}$, according to Ref.~\cite{Ganeshan2016largecosine}. 

Let us now place the system on a $L\times L$ torus. Just as $\mathcal{N}$, we find that $\mathcal{Z}$ is block diagonal: $\mathcal{Z}=\mathcal{Z}^{(2)}\oplus \mathcal{Z}^{(-4)}$, and $\sqrt{|\det \mathcal{Z}^{(k)}|}=|2k|^{L^2}$.\footnote{This can be computed explicitly by making use of the translation invariance and applying a Fourier transform.} One can further show, as the dimension of the subspace suggests, that $\mathcal{H}_{\rm eff}=\bigotimes_{i=1}^{L^2} \bC^{2}\otimes \bC^{4}$. We can choose a complete set of stabilizers for the subspace as $A_v^{(2)}$ and $A_v^{(-4)}$, for all $v$, along with the logical operators $\bar{Z}_2$ and $\bar{Z}_{-4}$.
This is redundant however, due to the relation $\prod_v A_v^{(2)}=\prod_v A_v^{(-4)}=1$, so that the number of independent $A_v^{(2)}$ operators is $L^2-1$. Since $(A_v^{(2)})^2=1$ in the code space, the number of states labeled by the eigenvalues of $A_v^{(2)}$ is $2^{L^2-1}$. Similarly, the number of states labeled by the $A_v^{(-4)}$ eigenvalues is $4^{L^2-1}$. Together with the logical operators, they completely describe the subspace $\mathcal{H}_{\rm eff}$.

At this point we add the vertex terms $H_A$ as a perturbation (so $U'\ll \sqrt{U\alpha}$). Because $A_{v}^{(2)}$ and $A_{v}^{(-4)}$ commute with the condensation terms, they act entirely within the subspace $\mathcal{H}_\mathrm{eff}$ as Pauli operators. Thus, in the spirit of degenerate perturbation theory, assuming $U'\ll \Delta\sim \sqrt{U\alpha}$, we can work within $\mathcal{H}_\mathrm{eff}$, with the following effective Hamiltonian:
\begin{equation}
    H_{\rm eff}=-U'\sum_v A_v^{(2)}-U'\sum_v A_v^{(-4)}.
\end{equation}
This selects the states with $A_v^{(2)}=A_v^{(-4)}=1$ as the true ground states with a gap $\sim U'$ to the excited states. The logical states remain degenerate, so the ground state degeneracy is $2\times 4=8$.

\subsubsection{Gapped Hamiltonian for $\U_2$}
We now repeat the analysis to construct a gapped model that hosts the excitations corresponding to the $\U_2$ anyon theory. We consider, in particular, the following Hamiltonian:
\begin{equation}
    H=H_2+H_0,
\end{equation}
with the quadratic perturbation $H_0$ given as before:
\begin{eqs}
    H_0=\alpha \sum_e[(w_e^{(2)})^2+(w_e^{(-4)})^2].
\end{eqs}
Due to the factorization of the Hilbert space, within the $\U_2$ subsystem, the Hamiltonian is identical to the one analyzed in Section~\ref{sec:gapping 1}, which has an energy gap and a 2-fold degenerate ground state subspace on a torus. On the other hand, there are no longer any stabilizer terms for the $\U_{-4}$ subsystem, and the Hamiltonian is instead purely quadratic.

Thus, our remaining task is to solve the quadratic Hamiltonian $H_0^{(-4)}=\alpha\sum_{e}(w_e^{(-4)})^2$ and show that it is gapped. 
Diagonalizing $H_0^{(-4)}$ amounts to finding all operators
$b$ that are linear combinations of $w_e^{(-4)}$ such that $[b, H_0^{(-4)}]=Eb$. Assuming $b$ is of the general form $b=\sum_e u_e w_e^{(-4)}$, then the equation $[b, H_0^{(-4)}]=Eb$ gives:
\begin{equation}
     -2i\alpha \sum_{e'}K_{ee'}u_{e'} = E u_e.
\end{equation}
Here, we have defined the matrix $K_{ee'}$ by $iK_{ee'}=[w_e, w_{e'}]$. If the modes with positive energy are indexed by $\ell$, the Hamiltonian can be written as $H_0^{(-4)}=\sum_\ell E_\ell b_\ell^\dag b_\ell+\text{const}$. With periodic boundary conditions, we find that the spectrum of $iK$ is gapped, and therefore, $H_0^{(-4)}$ is gapped. It is also evident that $H_0^{(-4)}$ has a unique ground state defined by $b_\ell=0$, for all $\ell$, on a torus. Thus, we can conclude that the ground state of the Hamiltonian is characterized by the anyons of the $\U_2$ CS theory.

\section{Topological codes from condensation in an $\bR$ gauge theory}\label{sec:condensation}

In this section, we present a general construction of topological CV stabilizer codes, for which the $K_{2,-4}$ stabilizer code of the previous section is a special case.
We first give a definition of topological stabilizer codes on CVs. We then introduce a stabilizer model for a $\bR$ gauge theory, which is the starting point of our constructions. The examples in Section~\ref{sec:Examples} are derived from the $\bR$ gauge theory through condensing various bosons, summarized in Section~\ref{sec:boson_condensation}. 

\subsection{Definition of topological stabilizer codes on CVs}\label{sec: def topological codes}

To define topological CV stabilizer codes, we consider families of CV systems embedded in some fixed spatial dimension $D$. For the codes that we construct, $D$ is 2. Each CV hosts displacement operators, or Pauli operators, as introduced in Eq.~\eqref{eq:displacements-def}. Stabilizer groups are groups of mutually commuting products of displacement operators. We define a topological CV stabilizer code as

\begin{definition}[topological CV stabilizer code]\label{def:topological_CV}
    A family of stabilizer groups $\cS_L$, for varying system sizes $L$, defines a topological CV stabilizer code, if the following two conditions are fulfilled:
    \begin{enumerate}
        \item $\cS_L$ can be locally generated -- There exists a system-size independent range $r\in\bR$ such that for each system size there exists a set of generators for which each generator acts nontrivially only on CVs contained in a ball of radius $r$.\footnote{For concreteness, we think of the ball as being defined with respect to the Euclidean metric on $\bR^D$.}
        \item $\cS_L$ is locally maximal -- The only geometrically local displacement operators that commute with all stabilizers are themselves stabilizers.
    \end{enumerate}
\end{definition}
The above definition is a straightforward generalization of the definition of topological qudit stabilizer codes in Ref.~\cite{Bombin2012universal}.
The codes presented in this work additionally have translation invariance, generalizing the stabilizer codes discussed in Ref.~\cite{Haah2018classification}.

The logical algebra of topological CV stabilizer codes can be qualitatively different from that of discrete-variable codes.
The main difference is that a CV code can encode continuous degrees of freedom or discrete degrees of freedom.
More specifically, the logical operators can be those of encoded qudits, quantum rotors, or CVs.\footnote{Formally, we can identify that the logical algebra -- the commutant of the stabilizer algebra -- decomposes into independent subfactors,
\begin{align}
    \cC(\cS) \simeq \cS\otimes \cL\qq{with} \cL = \bigotimes_{i = 1}^{N_d} \cM_{d_i}\bigotimes_{j=1}^{N_r} \bL(L^2(\U))\bigotimes_{k = 1}^{N_c}\bL(L^2(\bR)),
\end{align}
where $\cL$ is the algebra that acts transitively on the codespace,
$\cM_{d_i}$ denotes the algebra of $d_i\times d_i$ complex matrices, acting irreducibly on $\bC^{d_i}$, and $\bL(\cH)$ is the algebra of (bounded) linear operators on $\cH$.
We say that a code defined by $\cS$ encodes $N_d$ qudits with dimensions $\{d_i\}_i$, $N_r$ rotors, and $N_c$ CVs.}
Note that, by quantum rotor, we have in mind an infinite-dimensional quantum system as described, for example, in Ref.~\cite{Albert2017}.

Similar to stabilizer codes on qubits, there is a notion of code distance for stabilizer codes on CVs, which captures the resilience of the logical information to displacement errors. To motivate the definition, recall that the code distance for a qubit stabilizer code is the minimum weight of a Pauli operator that is able to perform a nontrivial logical gate. This definition of code distance needs to be refined for CV stabilizer codes. This is because noise in CV systems is typically modeled as a Gaussian distribution of displacements. Therefore, small displacements on all of the CVs is typical and to be expected. Simply minimizing the size of the support of the logical operator, does not give a clear indication that the logical information is robust to small displacements across the entire system. Thus, to give a more physically motivated definition of the code distance, we weight displacements according to their magnitude.  

More specifically, consider a generic displacement $P$ of the form:
\begin{align}
    P = \prod_i X_i^{s_i}Z_i^{t_i},
\end{align}
which is a product of displacements at CVs indexed by $i$. We define the \textit{CV weight} of $P$ as the length of the vector formed by $s_i$ and $t_i$:
\begin{align}
    |P| = \left( \sum_i s_i^2 + \sum_i t_i^2 \right)^{\frac12}.
\end{align}
This definition more appropriately assigns a greater weight to large displacements. We take the code distance of stabilizer code with stabilizer group $\mathcal{S}$ to be the minimum CV weight of a displacement operator that enacts a nontrivial logical operator,
\begin{align}
    d = \min_{P \in \mathcal{N}(\mathcal{S})\backslash \cS} (\abs{P}),
\end{align}
where the minimization is over all displacements that commute with the stabilizers but are not stabilizers themselves.

As we will see in the next section, some topological CV stabilizer codes have a vanishing code distance due to a continuous set of stabilizers.
These can spread out a logical operator to act with arbitrarily small displacements on an extensive number of modes.
In contrast, codes with a discrete set of stabilizer generators can admit a macroscopic scaling of code distance, as illustrated in various examples throughout the next section.

\subsection{Model for a $\bR$ gauge theory} \label{sec: R gauge theory model}
Our construction is based on condensing bosons within a lattice model for a $\bR$ gauge theory.
To define the model, we consider a system of CVs on edges of a square lattice. We denote the set of vertices, edges, and plaquettes by $V$, $E$, and $P$, respectively.
On each edge, we consider the operator algebra generated by the set of displacement operators $X$ and $Z$, introduced in Eq.~\eqref{eq:displacements-def}.

We define the $\bR$ gauge theory Hamiltonian $H_\bR$ as
\begin{align}\label{eq:H-R-gauge}
    H_{\bR} = -\sum_{v \in V} \int_{\bR}\dd{\varphi} A_v^{\varphi} - \sum_{p \in P} \int_{\bR}\dd{c} B_p^{c},
\end{align}
where the vertex terms $A_v^{\varphi}$ and plaquette terms $B_p^{c}$ are represented graphically as:
\begin{align}
    A_v^{\varphi} = \vcenter{\hbox{\includegraphics[scale=0.3]{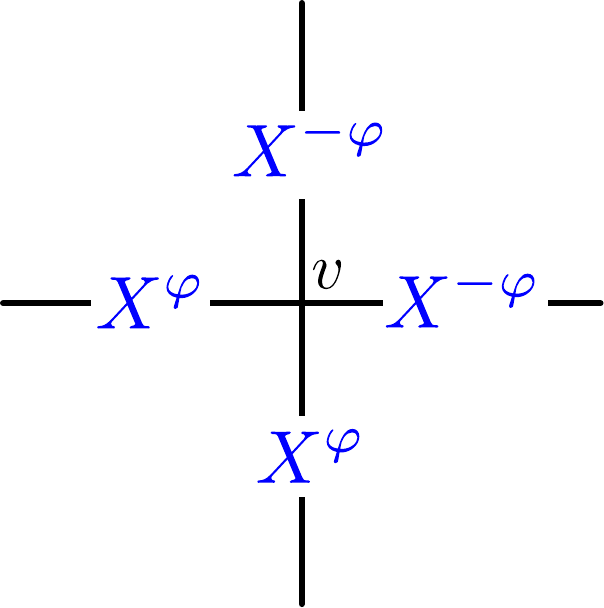}}}, \quad
    B_p^{c} = \vcenter{\hbox{\includegraphics[scale=0.3]{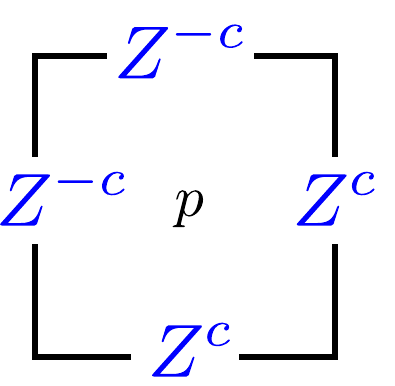}}}.
\end{align}    
Similar to a quantum double model for a finite group \cite{Kitaev2003fault-tolerant}, we can think of $A_v^{\varphi}$ as energetically imposing a Gauss's law,
while the integral over $B_p^{c}$ projects onto the flux-free subspace. This Hamiltonian can be interpreted as the natural generalization of the $\ZZ_n$ toric code to CVs.
Since all the terms in the Hamiltonian commute, we can identify the ground space as the space stabilized by the {stabilizer group}
\begin{align}\label{eq:R-stab-group}
    \cS_{\bR} = \langle \{A_v^{\varphi}, \, B_p^{c}\,|\, \varphi,c\in\bR \}_{v \in V,\,  p \in P}\rangle.
\end{align}
Note that the stabilizer group is continuous, since $\varphi$ and $c$ are valued in $\bR$. \\

\noindent\textbf{Remark:} We emphasize that the Hamiltonian $H_\bR$ for the $\bR$ gauge theory is unphysical, due to its delta-function-like interactions. More specifically, $A_v^\varphi$ and $B_p^c$ are complex exponentials of sums of quadratures. If the exponent evaluates to zero for a particular state, then the integrals over $\varphi,c \in \bR$ diverge. Otherwise, if the exponent evaluates to something nonzero, then the integrals vanish. Therefore, the interaction strength either diverges or is zero -- the couplings of the Hamiltonian are discontinuous and unbounded. Nonetheless, we find that the Hamiltonian $H_\bR$ offers valuable intuition. 
Later, in Section~\ref{sec:Examples}, we construct Hamiltonians that describe condensates of $H_\bR$.
For certain condensations, the resulting Hamiltonian does not have discontinuous interactions.
In particular, condensing two independent bosons yields a Hamiltonian that is a discrete sum of commuting displacement operators with bounded strength, similar to the Hamiltonian introduced in Section~\ref{sec: K24 hilbert space and code}.
Note that, independent of whether the Hamiltonian interactions are physical, the exact ground states of multi-mode GKP code Hamiltonians are not normalizable, as discussed in the remark in Section~\ref{sec: K24 hilbert space and code}.\\

\begin{figure}
    \centering
    \includegraphics[width=.8\textwidth]{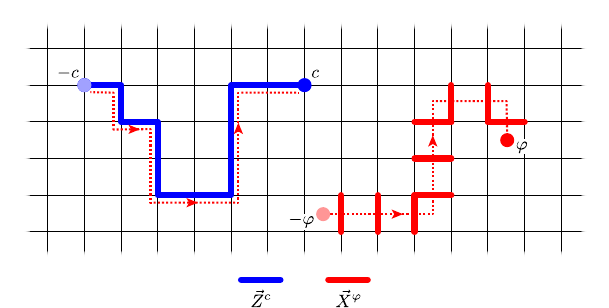}
    \caption{The topological excitations of the $\bR$ gauge theory Hamiltonian. The $\bR$ gauge theory Hamiltonian in Eq.~\eqref{eq:H-R-gauge} admits point-like excitations that are topological, in the sense that they cannot be created by operators supported solely in the vicinity of the excitation. The topological excitations are generated by the gauge charges, labeled by $c \in \bR$, and the gauge fluxes, labeled by $\varphi \in \bR$. The gauge charges and gauge fluxes are created by string operators on the direct lattice and dual lattice, respectively. Here, the notation $\vec{X}$ and $\vec{Z}$ denotes that the operator should be Hermitian conjugated if the orientation of the path (red dashed) points to the left or downward.}
    \label{fig:Rgauge}
\end{figure}

We next consider the local excitations of $H_\bR$. These correspond to violations of $A_v^{\varphi}$ and $B_p^{c}$ for some value of $\varphi,c \in \bR$. We refer to the violations of the $B_p^{c}$ operators as the gauge fluxes. These are created by string operators that are products of $X^{\pm\varphi}$ operators along a path in the dual lattice, as depicted in Fig.~\ref{fig:Rgauge}. The string operator create $\pm\varphi$ fluxes at the endpoints.  
Similarly, the gauge charges correspond to violations of the $A_v^{\varphi}$ operators. The charges are created by products of $Z^{\pm c}$ operators along a path in the primal lattice, as shown in Fig.~\ref{fig:Rgauge}.
The charges $\pm c$ are created at the endpoints of the path. 
Importantly, the string operators for both the fluxes and the charges commute with all of the Hamiltonian terms along their length. In this sense, they create point-like deconfined excitations. These excitations cannot be created by any operator restricted to an endpoint, so they belong to nontrivial superselection sectors. 
More generally, the local excitations of the Hamiltonian $H_\bR$ may be composites of charges and fluxes. We label a flux-charge composite by $(\varphi,c)$, with flux $\varphi \in \bR$ and charge $c \in \bR$. These exhaust all of the local excitations of the model, since $X$ and $Z$ generate all of the displacement operators -- any other excitation must be a superposition of these eigenstates.

As expected in a gauge theory, the charges and fluxes exhibit an Aharonov-Bohm effect -- moving a charge around a flux produces a phase. This can be seen explicitly from the commutation relations of the string operators. That is, exchanging the order in which a $Z_e^c$ and $X_e^\varphi$ is applied to the CV on edge $e$ produces the phase $e^{i\varphi c}$. More generally, the exchange statistics and braiding relations for a composite $(\varphi,c)$ of a flux and a charge are given in Section~\ref{sec:R-gauge-excitations-abstract}.

Since the string operators that create charges and fluxes commute with the Hamiltonian along the length of the string, the string operators acting along homologically trivial
loops correspond to products of terms in the stabilizer group.
In fact, the group of homologically trivial loop operators is exactly the stabilizer group of the ground space, see Eq.~\eqref{eq:R-stab-group}, and act as the identity on the ground space.
Additionally, on a surface with nontrivial homology, string operators that act along nontrivial loops do not commute with all other such string operators, but do commute with all stabilizers.
On a torus, for example, there exist two pairs of string operators, indexed by a real number, that obey the commutation relation 
\begin{align}
    \overline{Z}_j^b \overline{X}_i^a = e^{ia b\delta_{i,j}} \overline{X}_i^a \overline{Z}_j^b\qq{} \forall a,b\in\bR,
\end{align}
where $i,j=1,2$.
The ground space transforms irreducibly under the group generated by these operators, and thus we find that these string operators can be identified with displacement operators of two effective CVs, encoded in the ground space of $H_\bR$. The encoded CVs are not fault tolerant, however, in agreement with Refs.~\cite{Niset2009nogo,Vuillot2019toricGKP,Hanggli_2022}.
We also note that, for this CV stabilizer code, the code distance, as defined in Section~\ref{sec: def topological codes}, vanishes in the limit of a large system size. This is a consequence of the fact that the logical operators are continuously parameterized. A product of arbitrarily small displacements around a non-contractible path can implement a nontrivial logical operator.

\subsection{Algebraic description of deconfined excitations}\label{sec:R-gauge-excitations-abstract}

Let us now discuss the abstract properties of the deconfined excitations, without any explicit reverence to the lattice model -- although, these properties can be derived from the string operators in Fig.~\ref{fig:Rgauge}.
The fusion rules of the excitations are captured by the Abelian group 
\begin{align}
    \bR\times\bR = \{(\varphi,c)\;|\; \varphi,c\in\bR\}.
\end{align}
The group structure above says that the charge and flux of a pair of excitations fuse independently according to the additional of real numbers, 
\begin{align}
    (\varphi, c) + (\varphi',c') = (\varphi + \varphi',c+c').
\end{align} 

The exchange statistics of a generic flux-charge composite $(\varphi, c)$ is given by
\begin{align}\label{eq:R-gauge-spin}
    \theta\Big((\varphi, c)\Big) = e^{is(\varphi, c)}\qq{with}
    s(\varphi, c) = \varphi c \mod 2\pi,
\end{align}
A flux-charge composite $(\varphi, c)$ is a boson, i.e., has trivial self-exchange statistics, if and only if ${s(\varphi, c) = 0}$.
Note that the expression for $s$ in Eq.~\eqref{eq:R-gauge-spin} is symmetric with respect to exchanging the flux $\varphi$ and the charge $c$.
Hence, we can assume without loss of generality that $c\neq 0$ and deduce that every boson (up to flux-charge exchange) is of the form
\begin{align}\label{eq:R-gauge-boson-general}
    \left(\frac{2\pi}{a}, ak \right)\qq{with} a\in\bR, \, k\in\bZ.
\end{align} 
For a more in-depth discussion of the data defining an Abelian anyon model we refer the interested reader to Refs.~\cite{Ellison2022stabilizer, Wang2020Abeliananyons}.
In fact, the fusion group together with the exchange statistics $s$,  fully define an Abelian anyon theory. Nonetheless, it is convenient to further define the mutual braiding phase between two flux-charge composites as: 
\begin{align}\label{eq:R-gauge-braiding-bilinear}
    B\Big{(}(\varphi, c), (\varphi',c')\Big{)} = e^{ib((\varphi, c), (\varphi',c'))} \qq{with}  b \Big((\varphi, c), (\varphi',c') \Big) = \varphi c' + \varphi' c \mod 2\pi,
\end{align}

Later, we will consider subgroups of bosons. We say a subgroup of excitations is a subgroup of bosons if all of the fusion products of a generating set of bosons take the form in Eq.~\eqref{eq:R-gauge-boson-general}.
This is equivalent to saying that the {mutual braiding phase} between two flux-charge composites in the subgroup is
\begin{align}
    B\Big{(}(\varphi, c), (\varphi',c')\Big{)} = 1.
\end{align}

\subsection{Boson condensation}\label{sec:boson_condensation}
Thus far, we have described the deconfined excitations in a Hamiltonian model for a $\bR$ gauge theory.
In the following, we explain how to obtain new anyon models from {condensing} a subgroup of bosons in the parent $\bR$ gauge theory.
We first describe boson condensation on the abstract level with a generic bosonic subgroup.
Then we list all the condensations for which we construct lattice models in Section~\ref{sec:Examples}, demonstrating that different condensations can lead to qualitatively different anyon theories.

\subsubsection{Condensation in $\bR$ gauge theory}
On the abstract level, the procedure of condensation in an Abelian theory with fusion group $\cA$ involves three steps:
\begin{enumerate}
    \item Pick a subgroup of bosons $\cB\leq \cA$
    \item \textit{Confinement:} Decompose the elements in $\cA$ into two subsets: $\cA_{\cB}$, the subgroup of anyons that braid trivially with all bosons in $\cB$, and the subset of anyons that braid nontrivially with at least one boson in $\cB$.
    The anyons in $\cA_{\cB}$ are deconfined, while the latter set of excitations become {confined} after condensing $\cB$.
    \item \textit{Identification:} The set of deconfined excitations after condensation are $\cA_{\cB}$. Two elements in $\cA_{\cB}$ become {identified} if they differ by a boson in $\cB$.
    This leads to the set of anyons in the condensed theory being labeled by the quotient group $\cA_{\cB}/\cB$.
    As every element in $\cA_{\cB}$ braids trivially with $\cB$ the braiding data in the parent theory can be unambiguously lifted onto $\cA_{\cB}/\cB$.
\end{enumerate}
In the following, we list the different types of condensations based on the structure of $\cB$.
Since the bosons in an $\bR$ gauge theory are labeled by elements in $\bR\times\bZ$, see Eq.~\eqref{eq:R-gauge-boson-general}, $\cB$ must be a subgroup of $\bR\times\bZ$.\footnote{Note that although the bosons are labeled by $\bR \times \bZ$ their fusion is not described by that group. Two bosons can fuse to an excitation that is not a boson.}
A simplification occurs if $\cB$ contains a continuous part, i.e., $\bR\leq \cB$.
Since $\cB$ must be bosonic, the only continuous subgroups of bosons are pure charges or fluxes, see Eq.~\eqref{eq:R-gauge-braiding-bilinear}.
Hence, up to flux-charge exchange, $\cB$ must include a subgroup $\cB_f = \{(\varphi, 0)\;|\;\varphi\in\bR\}$.
Following the recipe above we find that $\cA_{\cB_f} = \cB_f$ and hence the condensate only has trivial excitations.
We can conclude that whenever $\cB$ admits a continuous factor, the condensate only has topologically trivial excitations and hence the ground space of any model implementing the condensate is one-dimensional and does not encode any logical information.

We obtain more interesting models by condensing discrete subgroups of bosons.
There are two classes of discrete condensates: $\cB\simeq\bZ$ and $\cB\simeq\bZ\times\bZ$.
Depending on the possible choice of generators for the discrete subfactors, we obtain qualitatively different condensates (accounting for flux-charge exchange).
We group the condensates according to the number of independent generators of $\cB$.
Angular brackets should be understood as denoting the set obtained from any integer combinations of elements in the brackets.
We construct exactly soluble lattice model for the following condensates:
\begin{enumerate}
    \item Single condensation, $\cB\simeq\bZ$:
    \begin{enumerate}
        \item Condensing pure fluxes,
        \begin{align}
        \cB = \left\langle \left(\frac{2\pi}{a}, 0\right) \right\rangle    
        \end{align}
        for $a\in\bR\backslash\{0\}$. The $\bR$ fluxes are compactified to $\U$ gauge fluxes (via identification) and
        the charges become discretized (via confinement). Together, they fuse according to $\U\times\bZ$.
        On a torus, the associated stabilizer code encodes two quantum rotors.

        \item\label{item:twisted-U1-condensation} Condensing a charge-flux composite,
        \begin{align}
            \cB =\left\langle \left(\frac{2\pi}{a}, a n\right) \right\rangle
        \end{align}
        for $a\in\bR\backslash\{0\}, \,n\in\bZ\backslash\{0\}$. The topological excitations that remain deconfined are nontrivial flux-charge composites resembling the excitations of a $\U_{2n}$ gauge theory.
        The set of all deconfined excitations decouples, both under fusion and braiding, into a finite chiral part $\bZ_{2n}$ and a continuous anti-chiral part that is isomorphic to $\bR$.
        On a torus, the associated stabilizer code encodes one qudit of dimension $2n$ and a single CV.
    \end{enumerate}

    \item Double condensation, $\cB\simeq\bZ\times\bZ$:
    \begin{enumerate}
        \item Condensing pure fluxes and charges,
        \begin{align}
           \cB = \left\langle \left(\frac{2\pi}{a}, 0\right) , (0,an)\right\rangle    
        \end{align}
        for $a\in\bR\backslash\{0\}$ and $n\in\bZ\backslash\{0\}$. 
        The set of inequivalent deconfined excitations after condensation is finite and forms the fusion group $\bZ_n\times\bZ_n$.
        The braiding reduces to the braiding of a $\bZ_n$ toric code, a topological $\bZ_n$ gauge theory.
        On a torus, the assocaited stabilizer code encodes two qudits of dimension $n$. If the pure flux is condensed first, then one obtains a $\U$ gauge theory; subsequently condensing an $an$ charge produces a $\bZ_n$ gauge theory.

        \item Condensing independent flux-charge composites of the form
        \begin{align}
            \cB = \left\langle \left(\frac{2\pi}{a},na\right), \left(-\frac{2\pi}{a}\sqrt{\frac{m}{n}}, a\sqrt{n m}\right)\right\rangle 
        \end{align}
        for $a\in\bR\backslash\{0\}$, $n,m\in\bZ\backslash\{0\}$. 
        This example can be viewed as a subsequent condensation from the example \ref{item:twisted-U1-condensation}. In this case, it is a boson in the continuous subfactor of excitations that is condensed.
        After condensation, the inequivalent topological excitations form two decoupled subfactors, a chiral $\bZ_{2n}$ factor and a $\bZ_{2m}$ factor with opposite chirality.
        The full condensed theory is equivalent to the emergent anyons in 
        $\U_{2n}\times \U_{-2m}$ CS theory (see Ref.~\cite{Cano2014bulkedge}).
        On a torus the associated stabilizer code encodes a qudit of dimension $2n$ and a qudit of dimension $2m$.
        \item More generally, we consider condensing a bosonic subgroup generated by flux-charge composites of the form
        \begin{align}
            \cB = \left\langle \left(\frac{2\pi}{a}, an_1 \right), \left( \frac{2\pi n_2}{a c_2}, ac_2 \right)\right\rangle,
        \end{align}
        defined from $a\in\bR\backslash\{0\}$ and $n_1,n_2,n'\in \bZ$ such that $n'^2>n_1n_2$.
        To simplify the notation, we have introduced the quantity $c_2=n' + \sqrt{{n'^2 -n_1n_2}}$.
        This condensate leads to an anyon theory of a non-chiral $\U\times\U$ CS theory with an even-valued $K$ matrix \cite{CSpaper}.
        On a torus, the associated stabilizer code encodes a qudit of dimension $d_1$ and a qudit of dimension $d_2$ such that $d_1d_2=4(n'^2-n_1n_2)$.
        The exact values for $d_1$ and $d_2$ depend on the choice of parameters $n_1,n_2,n'$.
    \end{enumerate}
\end{enumerate}
In the next section, we work out the details of the individual examples explicitly and show how to realize these condensates microscopically in exactly-solvable lattice models.
Important to our construction is the fact that the specific hopping terms have to be chosen carefully such that their commutation phase exactly agrees with the mutual braiding in the parent $\bR$ gauge theory.
In particular, the construction requires the exact commutation of all hopping terms associated to bosons with trivial mutual braiding.
We find such hopping terms for the condensates listed above. 
This includes theories that do not admit a Lagrangian subgroup and hence cannot be equipped with a fully gapped topological boundary, as for example the anyons in the code presented in Section~\ref{sec:example2-4}.

In all cases, the parameter $a$ that enters the definition of the bosonic subgroups does not affect the resulting theory formed by the deconfined excitations after condensation.
Different values of $a$ merely correspond to a rescaling of the local degrees of freedom.
Hence, we only consider the case $a=1$.

\section{Examples}
\label{sec:Examples}

In this section, we discuss in more detail examples of topological CV stabilizer models realized via boson condensation, starting from the model of an $\bR$ gauge theory in Section~\ref{sec: R gauge theory model}.
We recover all known classes of two-dimensional topological codes on CVs up to local unitary equivalence. (See Ref.~\cite{Victor2024tiger}, however, for a new class of topological codes on CVs that are outside of the Pauli stabilizer formalism.)
Furthermore, we identify models that realize anyon theories that are expected to be unrealizable in any topological stabilizer model on finite degrees of freedom. Throughout this section we use the notation of the $\bR$ gauge theory model introduced in the previous section.
All of the models in this section are defined on a square lattice with $V$ denoting the set of vertices, $E$ the set of edges and $P$ the set of plaquettes.

\subsection{$\U$ gauge theory}\label{sec:single-condensation-untwisted}
Our first example produces a $\U$ gauge theory. Consider the subgroup of bosons consisting of integer multiples of $2\pi$ fluxes:
\begin{align}
    \cB = \left\langle \left(2\pi, 0\right) \right\rangle.
\end{align}
The set of deconfined excitations after condensing $\cB$ is given by the solutions to the equation
\begin{align}
    b\left(\left(2\pi, 0\right), (x, y)\right) = 2\pi y \equiv  0 \pmod{2\pi},
\end{align}
which we identify with
\begin{align}
    \cA_{\cB} = \left\{ \left(x, n\right) \;|\; x\in\bR,n\in\bZ\right\}\simeq \bR\times \bZ.
\end{align}
Moreover, excitations are identified if they differ by a boson in $\cB$, so inequivalent excitations are in 1-1 correspondence with $\cB$ cosets $\cA_\cB/\cB$. This identification does not affect the $\bZ$ factor of $\cA_{\cB}$ but compactifies the $\bR$ part to $[0,2\pi)\simeq \U$.
We find that the topological excitations after condensation form the fusion group 
\begin{align}
    \faktor{\cA_{\cB}}{\cB} = \{(\varphi, m)\;|\; \varphi\in [0,2\pi), m\in\bZ\}\simeq \U\times \bZ .
\end{align}
Here, the $\bZ$ factor can be understood as charge excitations of a $\U$ gauge theory, while the $\U$ factor corresponds to the gauge fluxes.
The braiding phase, see Eq.~\eqref{eq:R-gauge-braiding-bilinear}, reduces to
\begin{align}\label{eq:single-condensation-untwisted-braiding}
    b((\varphi, m), (\varphi', m')) = \varphi m' + \varphi' m \mod 2\pi,
\end{align}
which agrees with the mutual braiding of the flux-charge composites in a $\U$ gauge theory.

\subsubsection{Lattice implementation -- homological rotor codes}
To implement the $2 \pi$ flux condensation at the level of the lattice, we start with the stabilizer group $\cS_{\bR}$ in Eq.~\eqref{eq:R-stab-group}. 
The $2 \pi$ fluxes are created by the single-CV displacement operators $X^{2\pi}$.
These operators generate the full group of string operators for bosons in $\cB$. Importantly, these hopping operators are mutually commuting. This means that they can be simultaneously measured, and thus they are valid hopping terms for stabilizing the states in which $\cB$ is condensed.
In a slight abuse of notation, we define the Abelian group of hopping terms for the condensed bosons as
\begin{align}
    \cB_{2\pi} = \langle \{X_e^{2\pi m} \, |\, m \in \ZZ  \}_{e\in E} \rangle .
\end{align}
To implement the condensation, we imagine measuring the generators of $\cB_{2\pi}$ (and assuming that the measurement outcome is $+1$).
Following the usual prescription for measurements in the stabilizer formalism, we first identify the centralizer of $\cB_{2\pi}$ over $\cS_{\bR}$, denoted $C_{\cS_{\bR}}(\cB_{2\pi})$. The centralizer captures the stabilizers that persist in the condensed model.
We find that the centralizer\footnote{That is, the stabilizers of the $\bR$ model that commute exactly with each element in $\cB_{2\pi}$.} is
\begin{align}
    C_{\cS_{\bR}}(\cB_{2\pi}) = \left\langle \{A_v^x, \, B_p^n \;|\; x\in \bR, \, n\in\bZ \}_{v\in V,\,  p\in P}\right\rangle. 
\end{align}
The stabilizer group of the $\cB$-condensed state is then generated by the centralizer and the hopping operators:
\begin{align}
\begin{split}
    \cS_{2\pi} = \left\langle C_{\cS_{\bR}}(\cB_{2\pi}), \cB_{2\pi}\right\rangle
    = \left\langle \{A_v^x, \, B_p^n, \, X_e^{2\pi m } \,  \;|\; x\in [0,2\pi), \, n,m\in\bZ \}_{v\in V, \, e\in E, \, p \in P}\right\rangle.
\end{split}
\end{align}

The logical operators of the $\cB$-condensed code are inherited from the logical operators of the parent $\bR$ gauge theory. In particular, they are the logical operators that belong to the commutant of $\cB_{2\pi}$.
Since $\cB_{2\pi}$ consists only of $X$ displacements, the logical operators composed of solely $X$ displacements remain logical operators after condensation.
For each nontrivial cocycle, i.e., non-contractible loop on the dual lattice $\gamma$, there exists an $X$-type logical operator of the parent code of the form\footnote{Here, we take $e \in \gamma$ to mean that the cocycle $\gamma$ evaluated on $e$ is nonzero.}
\begin{align}\label{eq:R-gauge-logicalX}
    \overline{X}_\gamma^a = \prod_{e\in\gamma} X_e^a \qcomma a\in\bR.
\end{align}
After condensation, $\overline{X}_\gamma^\theta$ and $\overline{X}_\gamma^{\theta+2\pi}$ become identified since $\overline{X}^{2\pi}_\gamma \in \cB_{2\pi}$.
Hence, the $X$-type logical operators are parameterized by $\theta \in (0,2 \pi]$ and are supported on nontrivial cocycles. 

Additionally, there are $Z$-type logical operators that remain after the condensation. Consider a nontrivial cycle, i.e., a non-contractible loop $\eta$ on the direct lattice, which has intersection number $I(\eta,\gamma)\in\ZZ$ with $\gamma$. 
Before condensation, there is an $\bR$-labeled set of $Z$-type logical operators along $\eta$ of the form
\begin{align}\label{eq:R-gauge-logicalZ}
    \overline{Z}_\eta^{r} = \prod_{e\in\eta} Z_e^r \qcomma r\in\bR.
\end{align}
However, only a $\bZ$-labeled subset of these operators remain logical operators -- i.e., commute with the hopping operators of $\cB_{2 \pi}$.
The logical commutation phase in the condensed model is inherited from the parent model, and exactly resembles the algebra of a quantum rotor:
\begin{align}
    \overline{Z}_\eta^{m} \overline{X}_\gamma^\theta = e^{i \theta m I(\eta,\gamma)} \overline{X}_\gamma^\theta \overline{Z}_\eta^{m},
\end{align}
where $m\in\bZ$, $\theta \in[0,2\pi)$. 
For each handle of the ambient manifold, there exists two cycle-cocycle pairs $(\eta_1, \gamma_1)$  and $(\eta_2,\gamma_2)$, such that $I(\eta_i, \gamma_j)=\delta_{ij}$, along which two independent subgroups of logical operators can be defined: $\{\overline{Z}_{\eta_j}^{m} \overline{X}_{\gamma_j}^\theta\;|\; \theta \in\ [0,2\pi), \, m\in \bZ, \, j=1,2\}$.
Each operator assigned to a pair only commutes nontrivially with operators of the same pair, e.g., $\overline{X}_{\eta_1}^\theta$ and $\overline{Z}_{\gamma_1}^m$.  
On a torus, for example,  $\cS_{2\pi}$ encodes $L^2(\U)^{\otimes 2}$; that is, two quantum rotors.

We remark that this construction recovers the homological rotor codes introduced in Ref.~\cite{vuillot2023homological}. In particular,  by enforcing the single CV stabilizer $X_e^{2\pi}$ on each edge, the operator algebra of displacements is reduced to the operator algebra of a single quantum rotor. In other words, the stabilizer $X_e^{2\pi}$ defines an encoding of a rotor in a CV. The stabilizers in $C_{\cS_{\bR}}(\cB_{2\pi})$ can then be viewed as those of a homological rotor code, concatenated with the encoding of the rotors into the CVs.
As acknowledged in Ref.~\cite{vuillot2023homological}, the distance of the homological rotor codes vanish in the thermodynamic limit.

The generators of the stabilizer group $\cS$ define a Hamiltonian for a $\U$ gauge theory. Explicitly, the Hamiltonian is
\begin{align}\label{eq:H11}
    H_{2\pi} = - \sum_{v\in V} \int_0^{2\pi}\dd{x} A_v^x - \sum_{p \in P} B_p - \sum_{e \in E} X_e^{2\pi} + \text{h.c.}.
\end{align}
This Hamiltonian is topological in the sense that the exact ground states are locally indistingushable and the dimension of the ground state space only depends on the homology group of the ambient manifold.
In contrast to qudit topological codes, the above Hamiltonian is gapless. Specifically, there exists local operators that create excitations of arbitrary small energy (e.g. $X_e^{\varepsilon}$ for arbitrarily small $\varepsilon\in\bR$).

\subsection{$\bZ_n$ gauge theory}

In this section, we consider the condensation of the bosonic subgroup 
\begin{align}
   \cB = \left\langle \left(2\pi, 0\right), \left(0,n \right)\right\rangle\qcomma n\in\bZ^+.
\end{align}
Condensing this subgroup can be viewed as first performing a condensation of a $2\pi$ flux, described in Section~\ref{sec:single-condensation-untwisted}, and then performing a second condensation of bosons generated by $(0,n)$. 
This perspective simplifies the analysis, as we can start with the deconfined excitations after the first condensation, which generated the group $\U\times\bZ$. 
The boson $(0,n)$ can be viewed as a pure charge in the $\bZ$ subfactor.
From the braiding in the $\U$ factor, see Eq.~\eqref{eq:single-condensation-untwisted-braiding}, we find that condensing this charge compactifies $\bZ\to\bZ_{n}$ and discretizes $\U\to\bZ_n$. 
Specifically, the set of deconfined excitations after the second condensation can be identified with
\begin{align}
    \faktor{\cA_{\cB}}{\cB} = \left\{\left( \frac{2\pi}{n}q,k\right) + \cB \;|\; q,k\in\{0,1,2,...,n-1\}\;\right\}\simeq \bZ_n\times\bZ_n.
\end{align}
The braiding reduces to
\begin{align}
    b\left(\left(\frac{2\pi}{n} q ,k\right),\left(\frac{2\pi}{n}q',k'\right)\right) = \frac{2\pi}{n} \left(qk'+q'k\right) \mod 2\pi,
\end{align}
which agrees exactly with the braiding of in a $\bZ_n$ gauge theory. 

\subsubsection{Lattice implementation -- Toric-GKP code}
Building on the lattice model in Section~\ref{sec:single-condensation-untwisted}, we extend the group of hopping terms by adding generators to $\cB_{2\pi}$ that serve as hopping terms of the charges $(0,n)$.
A valid choice for these hopping terms is given by the single-CV displacement operators $Z^{n}$.
This leads to the group of hopping terms 
\begin{align}
    \cB_{2\pi, n} = \langle \{X_e^{2\pi}, \, Z_e^{n} \}_{e\in E} \rangle.
\end{align}
Here, and unless otherwise specified, we take the angled brackets to imply that the group is generated by integer powers of the elements.  

To construct the stabilizer group of the condensed theory, we follow the same procedure as in Section~\ref{sec:single-condensation-untwisted} and determine the centralizer of $\cB_n$ over $\cS_{\bR}$.
We obtain
\begin{align}
    C_{S_\bR}(\cB_{2\pi,n}) = \left\langle \left\{A_v^{2\pi/n}\right\}_{v \in V}, \left\{B_p\right\}_{p \in P} \right\rangle
\end{align}
yielding a stabilizer group for the $\cB$-condensed theory
\begin{align}
    \cS_{2\pi,n} = \left\langle C_{S_\bR}(\cB_{2\pi,n}), \cB_{2\pi,n} \right\rangle
    = \left\langle \left\{ A_v^{2\pi/n} \right\}_{v\in V}, \left\{B_p\right\}_{p \in P}, \{X_e^{2\pi}, Z_e^{n}\}_{e \in E}  \right\rangle.
\end{align}

The logical operators are again a subset of those of the parent model.\footnote{Since condensing $\cB$ can be interpreted as a two-step process where $\cB$ is condensed first, the final logical operators are a subset of the logical operators from the first condensation.}
Since the stabilizer group decomposes into independent factors generated by $X$-type stabilizers and $Z$-type stabilizers, we can analyze the $X$-type and $Z$-type logical operators independently.
We find that for each nontrivial cocycle $\gamma$, there exists a family of $X$-type logical operators given by
\begin{align}
    \overline{X}^a_\gamma = \prod_{e\in\gamma} X_e^a \qcomma a\in \frac{2\pi}{n}\bZ.
\end{align}
Note that $\overline{X}^{2\pi}\in\cB_{2\pi,n}$.
Hence, this operator acts trivially on the subspace stabilized by $\cS_{2\pi,n}$.

The full algebra of operators on the code space is completed by including the $Z$-type logical operators.
Again, consider a nontrivial cycle $\eta$ which has intersection number $I(\eta,\gamma)$ mod $n$ with $\gamma$. 
On this cycle, we can define products of $Z$ operators,
\begin{align}
    \overline{Z}_\eta^q = \prod_{e\in\eta} Z_e^q \qcomma q\in \bZ,
\end{align}
each of which commutes with $\cS_{2\pi,n}$ and is a nontrivial logical operator in the parent model.
After condensation, the $Z$-type logical operator is order $n$, since $\overline{Z}_\eta^{n}\in \cS_{2\pi,n}$.
Additionally, there exists a basis of (co)cycles that can be grouped into pairs $(\gamma, \eta)$ such that the associated operators obey the commutation relations
\begin{align}
    \overline{Z}_{\eta}^{q 2\pi/n}\overline{X}_{\gamma}^{k} = e^{\frac{2\pi i kq}{n}I(\eta,\gamma)} \overline{X}_{\gamma}^{ k} \overline{Z}_{\eta}^{q 2\pi/n}.
\end{align}
This reproduces the commutation relations of generalized Pauli operators on qudits of dimension $n$.
We conclude that the stabilizer code defined by $\cS_{2\pi,n}$ encodes two qudits per handle.
On a torus, for example, $\cS_{2\pi,n}$ encodes a logical algebra of $\cM_{n}\otimes \cM_{n}$, acting faithfully on two qudits of dimension $n$.
In fact, the stabilizer group can be viewed as the stabilizer group of a {toric-GKP code} obtained from concatenating a single-CV GKP code, defined by the stabilizers $X_e^{2\pi}$ and $Z_e^{n}$, and a $\bZ_n$ toric code~\cite{Kitaev2003fault-tolerant}.
The toric-GKP code has a non-vanishing code distance, inherited from the $\bZ_n$ toric code. We want to remark that this property is ultimately due to the discrete set of deconfined excitations of the CV code.

We can further construct a stabilizer Hamiltonian that hosts the $\cB$-condensate in its ground state subspace,
\begin{align}
    H_{2\pi,n} = -\sum_{v\in V} A_v^{2\pi/n} - \sum_{p \in P} B_p - \sum_{e \in E}\left( Z_e^{n} + X_e^{2\pi}\right) + \text{h.c.}
\end{align}
In the subspace for which $X_e^{2\pi}=Z_e^{n} =1$, this is nothing other than the $\ZZ_n$ toric code Hamiltonian.

In contrast to topological stabilizer Hamiltonians on finite-dimensional systems, the Hamiltonian $H_{2\pi,n}$ is gapless, since local displacement operators can create excitations with arbitrarily small energy. 
This is reminiscent of GKP-code Hamiltonians~\cite{GKP2001} 
An important distinction to make about the spectrum of the Hamiltonian $H_{2\pi,n}$ is that all of the deconfined excitations are gapped, as it realizes a finite Abelian anyon theory, described by a $\bZ_n$ gauge theory.
Similar to the example from Section~\ref{sec:example2-4}, and the more general family presented in the next section, all gapless excitations are confined, since they do not commute with the hopping terms of the condensed bosons.
Following similar arguments to those presented in Section~\ref{sec: K24 gapping bulk}, a quadratic perturbation to $H_{2\pi,n}$ is sufficient to open a finite energy gap~\cite{Ganeshan2016largecosine}.

\subsection{$\U_{2n}$}\label{sec:single-condensation-twisted}

\begin{figure}
    \centering
    \includegraphics[width=0.5\linewidth]{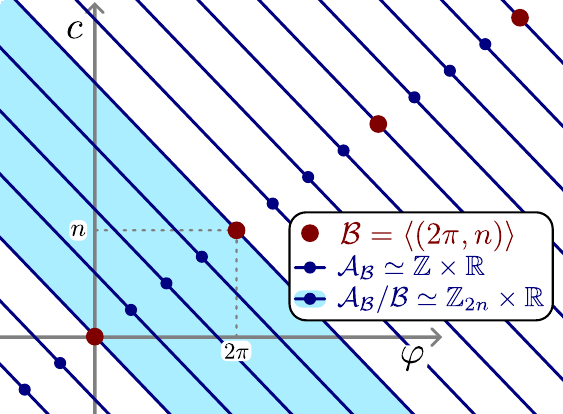}
    \caption{Effect of condensing a charge-flux composite in $\bR$ gauge theory. The $x$ axis corresponds to the flux label and the $y$ axis to the charge label of a deconfined excitation in the parent $\bR$ theory.
    The deconfined excitations can be identified with a $\bZ\times \bR$ subgroup (indicated with dark blue lines), after condensing a $\bZ$ subgroup of bosons (indicated with red dots).
    Additionally, condensation compactifies the space by adding equivalences among the excitations.
    In the case shown here, this leads to an effective description within the cylinder-shaped light blue region.
    The discrete part of the deconfined excitations (indicated with dark blue dots) gets compactified to a cyclic group while the continuous part is unaffected by the compactification since the $\bR$ parameter runs along the cylinder.
    }
    \label{fig:twisted_condensation}
\end{figure}

Next, we condense bosons in the $\bR$ gauge theory that are flux-charge composites. This leads to topological CV stabilizer codes that are beyond those in the current literature. In particular, let us consider a subgroup of bosons
\begin{align}\label{eq:composite_boson_single_condensation}
    \cB = \left\langle \left(2\pi,n\right) \right\rangle\qcomma n\in\bZ\backslash\{0\}.
\end{align}
Note that up to rescaling of the flux and charge variables\footnote{Concretely, given a constant $s\in\bR$ we can rescale $(\varphi,c) \mapsto (\varphi s, c/s)$ without changing the data of the topological excitations. In the lattice model this symmetry transformation can be implemented via an on-site \textit{squeezing} operation.} the $\U_2$ example discussed in Section~\ref{sec:example2-4} is built from condensing a subgroup of that form for $n=1$.

The set of deconfined excitations after condensation is given by the set of solutions to
\begin{align}
    b\left(\left(2\pi, n\right), (x, y)\right) = 2\pi y + n x \equiv 0 \pmod{2\pi},
\end{align}
which we identify with
\begin{align}\label{eq:def_AB_twistedcondensation}
\cA_{\cB} = \left\{(\varphi, k - \varphi n/2\pi) \;|\; \varphi\in \bR, k\in\bZ \right\} \simeq \bR\times \bZ.
\end{align}
We find that condensing $\cB$ discretizes the space of excitations along one direction which is illustrated in Fig.~\ref{fig:twisted_condensation}.
Additionally, specific combinations of charge-flux composites remain deconfined after condensation.
We show in this section that the identification of excitations by condensing $\cB$ results in a finite subgroup of deconfined excitations described by the anyons in a $\U_{2n}$ gauge theory~\cite{wen2004quantum,Cano2014bulkedge}.

The inequivalent excitations correspond to cosets with respect to $\cB$.
Since $\cB$ is isomorphic to $\bZ$, we expect $\cA_{\cB}\simeq \bR\times \bZ$ to become compactified along one direction.
In fact, we find that $(\pi/n, 1/2)$ generates a finite subgroup of $\cA_{\cB}/\cB$ of the order $2n$.
The group of cosets has another factor, formed by cosets represented by $(\alpha, -\alpha n/2\pi)$ for $\alpha\in\bR$.
There is no value $\Tilde{\alpha}\neq 0$ such that $(\Tilde{\alpha}, - \Tilde{\alpha} n/2\pi)\in \cB$.
Taken together, the group of deconfined excitations after condensing $\cB$ is given by
\begin{align}\label{eq:1condensation-twisted-anyons}
    \faktor{\cA_{\cB}}{\cB} = \left\{\left(q(\pi/n, 1/2) + (\alpha, -\alpha n/2\pi) \right) + \cB\;|\; q\in\bZ_{2n},\alpha\in\bR\right\} \simeq \bZ_{2n}\times\bR.
\end{align}
We give a rigorous proof of the isomorphism sketched in the text above in Appendix~\ref{app:twisted-condensation-group}.
In the following, we change the basis and label an element in $\cA_{\cB}/\cB$ by $[q,\alpha] := (q \pi/n + \alpha, q/2 + \alpha n/2\pi) + \cB$ with $q\in\bZ_{2n}$, $\alpha\in\bR$.
In this basis the braiding of the deconfined excitations after condensation reduces to
\begin{align}\label{eq:single-condensation-twisted-braiding}
    b([q,\alpha], [q',\alpha']) = (qq' - \alpha\alpha')\pi/n \mod 2\pi.
\end{align}
Each term only involves $q$ and $q'$ or $\alpha$ and $\alpha'$.
It follows that the group of deconfined excitations factorizes fully into a finite and an infinite subgroup.
In particular, the subtheory generated by $[1,0]$ forms a $\U_{2n}$ anyon theory.
However, we find that the infinite subfactor includes a finite subtheory with opposite chirality due to the minus sign entering Eq.~\eqref{eq:single-condensation-twisted-braiding}.
The cancellation can be seen more explicitly when a second boson is condensed, see Sections~\ref{sec:double-cond-2n-2m} and \ref{sec:general_double_condensation}.

\subsubsection{Lattice implementation}

In order to define an exactly soluble lattice model for the single condensation of a flux-charge composite, 
we have to find a suitable set of hopping operators which generate the string operators for the bosons that we want to condense in the $\bR$ gauge theory. Specifically, we want mutually commuting operators that form basis of all operators that create and move the condensed bosons when applied to a ground state of $H_{\bR}$.

To specify an appropriate set of hopping operators for the bosons, we first define a set of seven-body displacement operators labeled by a flux-charge composite ($\varphi,c$), with $\varphi,c \in \bR$. More specifically, on each edge $e$, we define the operators $\{C_e(\varphi, c)\;|\; \varphi, a\in\bR\}_{e \in E}$ with
\begin{align}\label{eq:def_Wehoppingterms}
    C_e(\varphi, c) = \vcenter{\hbox{\includegraphics[scale=0.3]{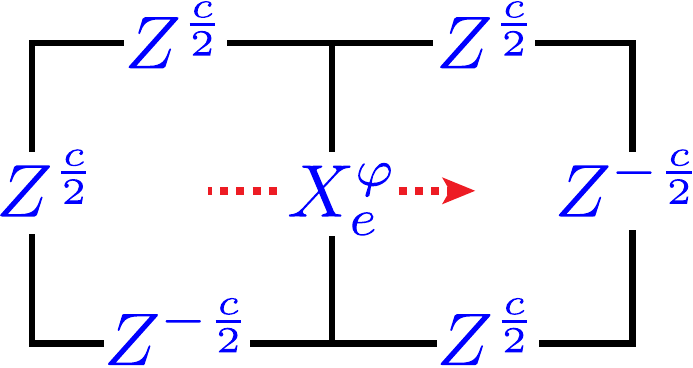}}}\,,\quad \vcenter{\hbox{\includegraphics[scale=0.3]{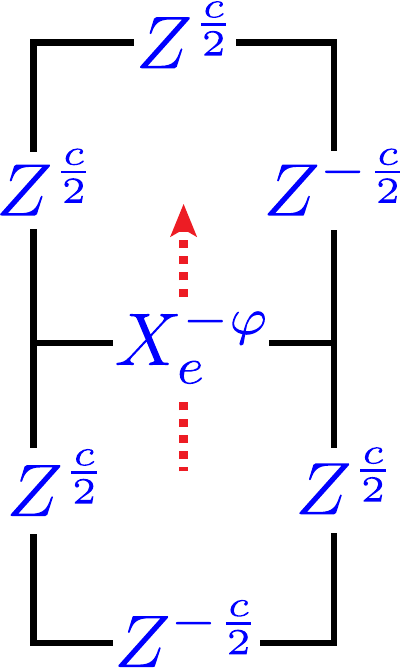}}}.
\end{align}
When these operators act on the ground space of $H_{\bR}$, they create flux-charge composites on adjacent sites with the charge bound to the south east of the flux,
\begin{align}
   \vcenter{\hbox{\includegraphics[scale=0.3]{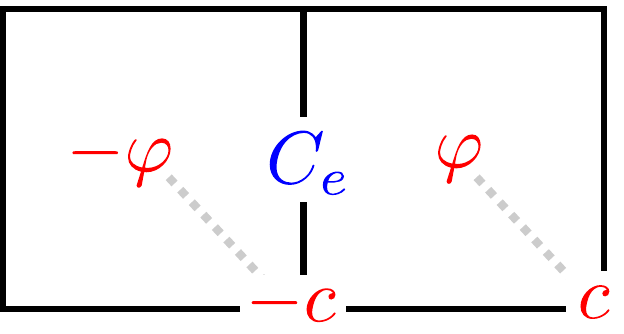}}}\,,\qquad \vcenter{\hbox{\includegraphics[scale=0.3]{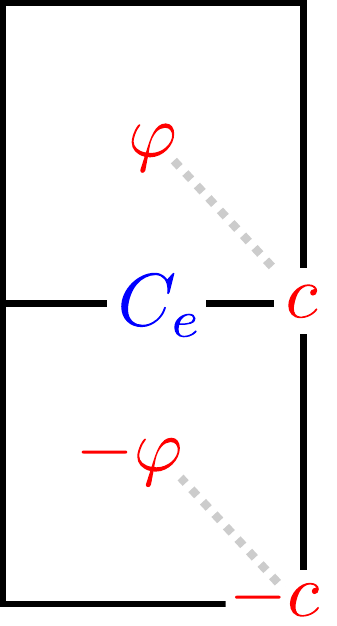}}}.
\end{align}
In the following, we refer to the above operators as $(\varphi,c)$-hopping operators since they generate  string operators for the excitations of $H_{\bR}$. 
In fact, these operators form a basis of all linear operators on $\cH$, as argued in Appendix~\ref{app:twisted_hoppingterms}.

For a detailed treatment of the algebra generated by the $C_e(\varphi, c)$ operators, we refer to Appendix~\ref{app:twisted_hoppingterms}. However, here we note their commutation relations.
Consider two different operators $C_e(\varphi, c)$ and $C_{e'}(\varphi',c')$.
Since they are a tensor product of displacement operators, they commute up to a phase
\begin{align}\label{eq:twisted-hopping-commutation}
    C_e(\varphi, c)C_{e'}(\varphi', c') = \psi_{e,e'}((\varphi, c), (\varphi',c'))C_{e'}(\varphi', c')C_e(\varphi, c).
\end{align}
A straightforward calculation shows that $\psi_{e,e'}((\varphi, c), (\varphi',c'))$ takes values
\begin{align}
    1 \qq{or} e^{\pm i(\varphi c' + \varphi'c)/2},
\end{align}
depending on the relative positions of $e$ and $e'$.
We find that the commutation phase agrees, up to a factor of $2$, with the braiding phase of the underlying excitations of $H_{\bR}$, see Eq.\,\eqref{eq:R-gauge-braiding-bilinear}.
This property is essential for the operators to commute exactly when the bosons that are condensed braid trivially i.e. have a braiding phase of $0$ and will be used explicitly \ref{sec:double-cond-2n-2m}, building on the model we present in the following.
We want to remark that these operators share similarities with hopping operators presented in Ref.~\cite{Haah2023invertible}. In that context, they were constructed to generate an {invertible subalgebra}. We think of the hopping operators used in our construction in the condensation of charge-flux composites as giving rise to a CV generalization of this type of algebra.

To define the lattice model for the $\cB$-condensate, consider the subgroup
\begin{align}
    \cB_n = \left\langle \left\{C_e\left(2\pi, n \right)\right\}_e\right\rangle \qcomma n\in\bZ\backslash \{0\}.
\end{align}
Evaluating $\psi_{e,e'}$ on the above subgroup, we find that all generators commute and hence generate a commutative subalgebra of linear operators on $\cH$.
To construct the lattice model for the condensate we first determine the centralizer of $\cB_n$ over $\cS_{\bR}$,
\begin{align}
    C_{\cS_{\bR}}(\cB_n) = \left\langle \left\{S_v(\varphi, k - n\varphi/2\pi)\;|\; \varphi\in\bR,k\in\bZ \right\}_{v \in V} \right\rangle,
\end{align}
where
\begin{align}\label{eq:composite-R-stabilizer}
    S_v(\varphi, c) = \vcenter{\hbox{\includegraphics[scale=0.3]{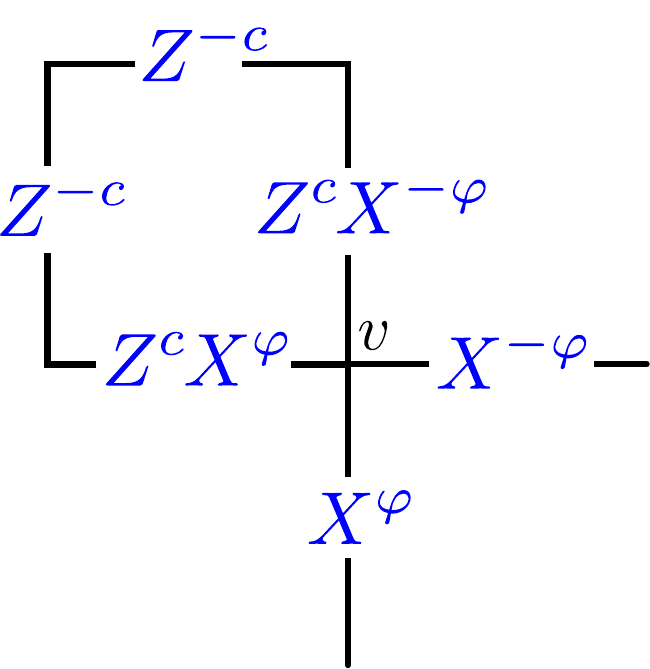}}} \in\cS_{\bR}.
\end{align}
Together with $\cB_n$, this yields a topological stabilizer group
\begin{align}
    \cS_n = \left\langle C_{\cS_{\bR}}(\cB_n),  \cB_n\right\rangle
    = \left\langle \left\{S_v(\varphi, k - n\varphi/2\pi)\;|\; \varphi\in\bR,k\in\bZ\right\}_{v\in V}, \left\{C_e\left(2\pi, n \right) \right\}_{e\in E} \right\rangle.
\end{align}

The logical operators are inherited from the parent model $\cS_{\bR}$.
In particular, they correspond to the logical operators of $\cS_{\bR}$ that also lie in the commutant of $\cB_n$.
The commutation relation of logical operators, generated by $C_e(\varphi, a)$ along cocycles is exactly the braiding phase of the excitations created by elements in $\cB_n$ and the excitations that are transported along the cocycle by the logical operator.
Hence, we can use the analysis leading to Eq.~\eqref{eq:1condensation-twisted-anyons} and find that for each nontrivial cocycle $\gamma$ there is a $\bZ_{2n}\times \bR$-labeled set of logical operators
\begin{align}
    \left\{\overline{W}_{\gamma}(q, \alpha)\;|\; q\in\bZ_{2n}, \alpha\in\bR \right\},
\end{align}
where $\overline{W}_{\gamma}(q, \alpha)$ is a string operator of the $\bR$ gauge theory model that transports a $[q,\alpha]$ flux-charge composite around a nontrivial cocycle.
Two logical operators associated to cohomologous cocycles differ by an element in $\cS$ and therefore commute.
Additionally, for a fixed $\gamma$, they form a representation of $\bZ_{2n}\times \bR$,
\begin{align}
    \overline{W}_{\gamma}(q, \alpha) \overline{W}_{\gamma}(q', \alpha') = \overline{W}_{\gamma}(q+q', \alpha+\alpha'),
\end{align}
where the $q$ values are added modulo $2n$ and $\alpha$ values are added as real numbers.
Moreover, given two inequivalent cocycles that have a nonzero intersection number, the commutation between the associated logical operators factorizes fully over the discrete $\bZ_{2n}$ factor and the continuous $\bR$ factor.
For example, on a torus, with the two inequivalent generating (co)cycles labeled as $\gamma$ and $\eta$, we find
\begin{align}
    \overline{W}_{\gamma}(q, \alpha) \overline{W}_{\eta}(q', \alpha') = e^{i (qq' \pi/n - \alpha\alpha')} \overline{W}_{\eta}(q', \alpha')\overline{W}_{\gamma}(q, \alpha).
\end{align}
In other words, $\cS_n$ encodes a qudit of dimension $2n$ and a CV.

Note that the code distance vanishes, due to the continuous deconfined excitations. However, the discrete excitations decouple from the continuous excitations, due to the factorization of the algebra of local operators, as demonstrated in Section~\ref{sec: K24 splitting of algebra} and Appendix~\ref{app:twisted_hoppingterms}. Interestingly, this implies that the logical qudit itself has macroscopic code distance. Therefore, this can be treated as a subsystem code with an extensive distance for the logical qudit.

The associated stabilizer Hamiltonian is
\begin{align}
\begin{split}
    H_n =& -\sum_v \left( \int_{\bR}\dd{\alpha} S_v(\alpha, -\alpha n/2\pi) + S_v(\pi/n,1/2)\right) - \sum_e C_e(2\pi, n)
    + \text{h.c.}
\end{split}
\end{align}
The ground space of $H_n$ is the mutual $+1$ eigenspace of all stabilizers in $\cS_n$.
Since $\cS_n$ is a topological stabilizer group, $H_n$ is topologically ordered in the same sense as $H_{2\pi}$, see Eq.~\eqref{eq:H11}.

In the following, we briefly comment on the spectrum of $H_n$ when the condensation is enforced exactly, i.e., we consider the Hamiltonian
\begin{align}
\begin{split}
    H'_n(J) =& -\sum_v \left( \int_{\bR}\dd{\alpha} S_v(\alpha, -\alpha n/2\pi) + S_v(\pi/n,1/2)\right) - J\sum_e C_e(2\pi, n)
    + \text{h.c.},
\end{split}
\end{align}
in the $J\to\infty$ limit.
It has two kinds of topological excitations. The first are {gapped} and discrete. These excitations can be created by integer powers of short string operators of the form
\begin{align}\label{eq:single-condensation-shortstrings-finite}
    W_e^{(2n)} = \vcenter{\hbox{\includegraphics[scale=0.3]{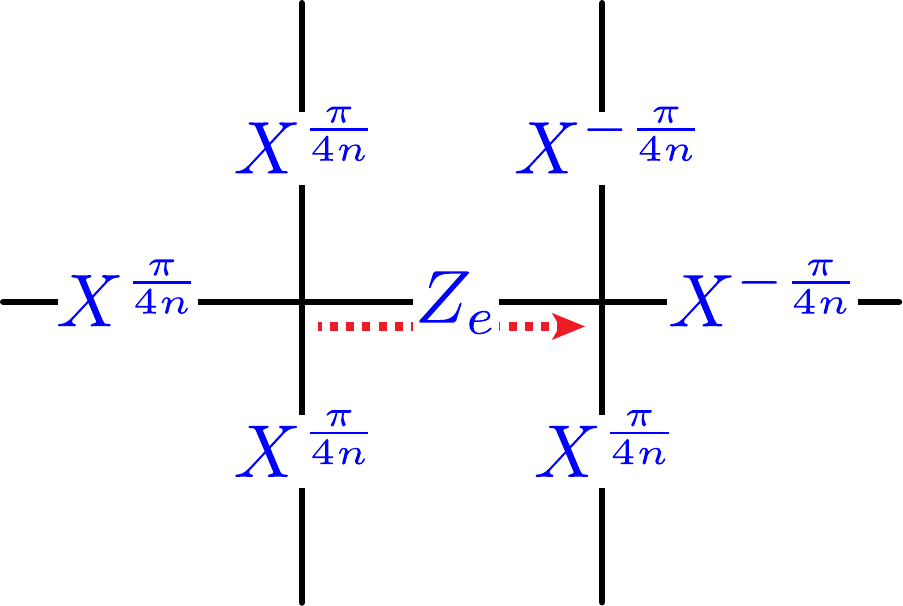}}}\,,\quad \vcenter{\hbox{\includegraphics[scale=0.3]{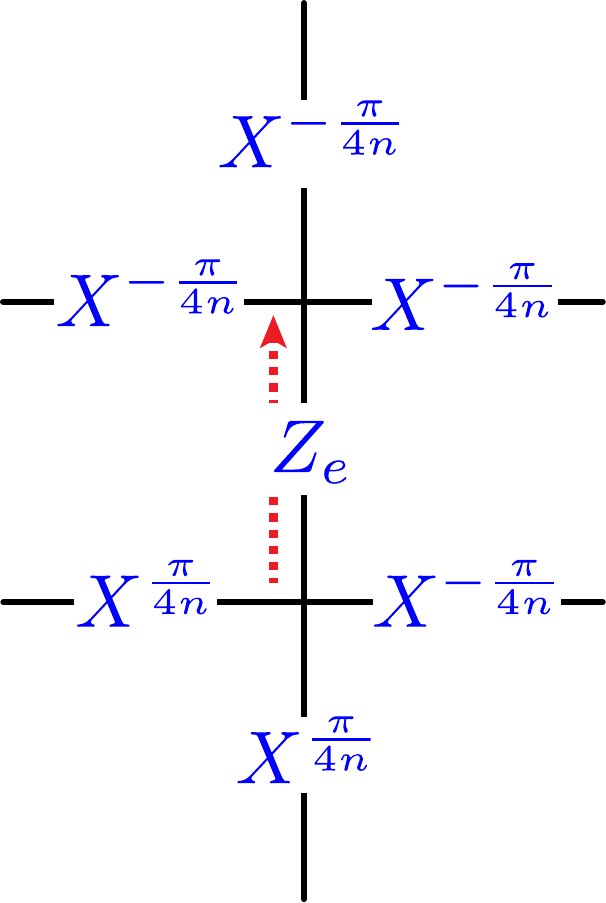}}}.
\end{align}
Each of the eigenstates that are obtained by applying products of these operators has a finite energy of 2 or more.
Moreover, applying these operators along a cycle leads to an operator that acts within the ground space, i.e., a logical operator.

Additionally, there is a {gapless}, continuous, set of topological excitations that can be created by products of operators $(W_e^{(2n)})^\alpha$, for $\alpha \in \bR \backslash\ZZ$.
The energy of the eigenstates obtained by applying one of these operators on the ground space is a continuous function in $\alpha$.
Since $\alpha$ can take any value in $\bR$ the energy can be arbitrarily small. This means $H_n$ has eigenstates of arbitrarily small energy.
However, the spectrum in the condensed subspace fully decouples into a gapped and a gapless part. In particular, there is no combination of gapless excitations that lead to gapped topological one.
Considering finite $J$, i.e., also allowing operators that do not commute with the condensation terms $C_e(2\pi, n)$, the Hamiltonian has additional confined excitations that can be created arbitrarily with local displacement operators.
Taking products of these operators along a path violates condensation terms along that path. In this sense they are confined excitations.

In the above model, we observe a similar factorization of local operators as in the model discussed in Section~\ref{sec: K24 splitting of algebra}.
We expect that adding suitable quadratic perturbations, derived from the factorization, leads to a gapped Hamiltonian whose ground space properties agree with that of $H_n$.
Similar to how a controllably-solvable model for a $\U_2$ theory was derived in Section~\ref{sec:example2-4}, this Hamiltonian presumably gives rise to controllably-solvable models for $\U_{2n}$ theories.

\subsection{$\U_{2n}\times \U_{-2m}$}\label{sec:double-cond-2n-2m}

In this section, we consider the condensation of the subgroup of bosons
\begin{align}
    \cB = \left\langle \left(2\pi,n\right), \left(-2\pi\sqrt{m/n}, \sqrt{n m}\right)\right\rangle.
\end{align}
This can be viewed as condensing a boson within the continuous part of topological excitation in the condensate described in the previous section, see Eq.~\eqref{eq:composite_boson_single_condensation}.

The group of deconfined excitations is defined by the solutions to the set of equations
\begin{subequations}
\begin{align}
    b\left(\left(2\pi,n\right),(x,y) \right) =& 2\pi y  + n x \mod 2\pi = 0 ,\\
    b\left( \left(-2\pi\sqrt{m/n}, \sqrt{n m}\right), (x,y) \right) =& -2\pi\sqrt{m/n} y + \sqrt{n m}x \mod 2\pi = 0.
\end{align}
\end{subequations}
We partially solved this set of equations in Section~\ref{sec:single-condensation-twisted}.
Building on this, we find that the group of deconfined excitations after condensing $\cB$ is
\begin{align}
    \cA_{\cB} = \left\{ \left(2\pi \left(\frac{k_1}{2n} + \frac{k_2}{2\sqrt{nm}} \right), \left(k_1 - \sqrt{\frac{n}{m}}k_2\right)/2\right) \;|\;  k_1,k_2\in\bZ\right\}\simeq \bZ\times\bZ.
\end{align}
For convenience, we introduce the notation 
\begin{align}\label{eq:ex_22_deconfined_label}
    [k_1,k_2]:= \left(2\pi(k_1/2n + k_2/2\sqrt{nm}), (k_1 - \sqrt{n/m}k_2)/2 \right).
\end{align}
The inequivalent excitations are given by cosets with respect to $\cB$.
Taking the quotient with respect to this group of condensed bosons yields a finite set of topologically nontrivial excitations,
\begin{align}
    \faktor{\cA_{\cB}}{\cB} \simeq \bZ_{2n}\times \bZ_{2m}.
\end{align}
The first factor can be identified with the factor of $\ZZ_{2n}$ in Eq.~\eqref{eq:1condensation-twisted-anyons}, and is generated by the coset with representative $[1,0]$. The second factor is generated by the coset with representative $[0,1]$.
For the remainder of this section we label each coset by one of its representatives $[k_1,k_2]$, for $k_1\in\bZ_{2n}$, $k_2\in\bZ_{2m}$, making a slight abuse of notation.

The braiding of the excitations in the condensate is given by
\begin{align}
    b\left( [k_1,k_2], [k_1',k_2']\right) = \pi\left(\frac{k_1k_1'}{2n} - \frac{k_2k_2'}{2m}\right) \mod 2\pi,
\end{align}
which agrees with the data of the anyons in $\U_{2n}\times \U_{-2m}$ CS theory.

\subsubsection{Lattice implementation}
We build on the construction of the topological stabilizer group $\cS_{n}$ from Section~\ref{sec:single-condensation-twisted} and consider the group of hopping terms
\begin{align}
    \cB_{n,m} = \left\langle \left\{ C_e\left(2\pi, n \right), C_e\left(-2\pi\sqrt{m/n}, \sqrt{nm} \right)\right\}_{e \in E}\right\rangle,
\end{align}
$n,m\in\bZ\backslash\{0\}$.
The operators labeled with $C_e$ are defined in Eq.~\eqref{eq:def_Wehoppingterms}.
It can be seen, by evaluating $\psi$ defined in Eq.~\eqref{eq:twisted-hopping-commutation}, that all the operators in this group commute. This is a nontrivial condition to satisfy and justifies the use of the hopping terms in Eq.~\eqref{eq:def_Wehoppingterms}.

In order to obtain a topological stabilizer group, we take the centralizer of $\cB_{n,m}$ over $\cS_{\bR}$,
\begin{align}
\begin{split}
    C_{\cS_{\bR}}(\cB_{n,m}) = \left\langle \Big\{ S_v\left([k_1,k_2] \right) \;\big|\; k_1,k_2\in\bZ\Big\}_{v \in V}
    \right\rangle,
\end{split}
\end{align}
where we have used the labeling from Eq.~\eqref{eq:ex_22_deconfined_label}.
This leads to the stabilizer group of the $\cB$-condensate
\begin{subequations}
\begin{align}
    \cS_{n,m} =&
    \left\langle C_{\cS_{\bR}}(\cB_{n,m}), \cB_{n,m} \right\rangle \\
            =& \Big\langle \left\{S_v([k_1,k_2]) \;\big|\; k_1\in\bZ_{2n}, k_2\in\bZ_{2m}\right\}_{v\in V},
            \left\{ C_e\left(2\pi, n \right), C_e\left(-2\pi\sqrt{m/n}, \sqrt{nm} \right)\right\}_{e \in E} \Big\rangle.
\end{align}    
\end{subequations}
Here, we have identified that $S_v([2n,0])$ and $S_v([0,2m])$ can be written as products of hopping terms in $\cB_{n,m}$.

The logical operators are again inherited from the parent model, and are those that commute with the elements of $\cB_{n,m}$.
Following the preceding analysis, we can conclude that, for each nontrivial cocycle $\gamma$, there exists a $\bZ_{2n}\times \bZ_{2m}$-labeled set of inequivalent logical operators,
\begin{align}
    \left\{\overline{W}_\gamma(k_1,k_2) \;|\; k_1\in\bZ_{2n},k_2\in\bZ_{2m} \right\},
\end{align}
each of which can be thought of as moving a deconfined excitation around a nontrivial cocycle.
Two such operators, with the same argument $k_1,k_2$ defined along cohomologous cocycles differ exactly by an element in $\cS_{n,m}$.
Additionally, up to elements in $\cS_{n,m}$, they represent the fusion group of the doubly condensed model for each cocycle,
\begin{align}
    \overline{W}_\gamma(k_1,k_2)\overline{W}_\gamma(k_1',k_2') = \overline{W}_\gamma(k_1+k_1',k_2+k_2'),
\end{align}
where the first argument is summed modulo $\bZ_{2n}$ and the second argument is summed modulo $\bZ_{2m}$. 

The logical operators associated to inequivalent cocycles generically do not commute.
We find that they exactly represent the anyon model defining the double $\cB$-condensate.
More specifically, the algebra generated by inequivalent logical operators splits into two independent factors, described by the anyon theory of $\U_{2n}\times \U_{-2m}$ CS theory.
On a torus, for example, with the generating cocycles $\gamma$ and $\eta$, the associated logical operators have the following commutation relation 
\begin{align}
    \overline{W}_{\gamma}(k_1, k_2) \overline{W}_{\eta}(k_1', k_2') = e^{i \pi (k_1k_1'/n - k_2k_2'/m)} \overline{W}_{\eta}(k_1', k_2') \overline{W}_{\gamma}(k_1, k_2).
\end{align}
Equivalently, they generate a logical algebra isomorphic to that of a $2n$-dimensional qudit and a $2m$-dimensional qudit. Therefore, the code defined by $\cS_{n,m}$ encodes two qudits, one with dimension $2n$, and one with dimension $2m$.
On a torus, this leads to a codespace dimension of $4nm$. 
We can easily pick integers $n,m$ such that this is not a square number.
In contrast, all known topological codes on qudits have a code space dimension on the torus that is a square number~\cite{Ellison2022stabilizer}, indicating that our construction goes beyond codes on finite-dimensional systems.
Given that the deconfined excitations are all discrete, the code distance is extensive in the system size. This follows from the same observation as in Section~\ref{sec: K24 splitting of algebra}.

The associated stabilizer Hamiltonian of $\cS_{n,m}$ is
\begin{align}
\begin{split}
    H_{n,m} =& -\sum_{v\in V} \left(S_v([1,0]) + S_v([0,1])\right)\\
    &- \sum_{e \in E} \left( C_e\left(2\pi, cn \right) +  C_e\left(-2\pi\sqrt{\frac{m}{n}}, \sqrt{nm} \right)\right)+ \text{h.c.}.
\end{split}
\end{align}
Since $\cS_{n,m}$ is a topological stabilizer group,  the above Hamiltonian is topologically ordered, and has a finite-dimensional ground space. Moreover, the Hamiltonian does not involve integrals over displacement operators with arbitrarily large strengths. Instead, only a finite set of displacement operators appear for each site.

The deconfined excitations are all gapped and can be created in pairs by short string operators of the form
\begin{subequations}
    \label{eq:twisted-double-condensation-short-strings}
    \begin{align}
    W_e^{(2n)} &= \vcenter{\hbox{\includegraphics[scale=0.3]{generalWh.pdf}}}\,,\quad \vcenter{\hbox{\includegraphics[scale=0.3]{generalWv.pdf}}}, \\
    W_e^{(-2m)} &= \vcenter{\hbox{\includegraphics[scale=0.3]{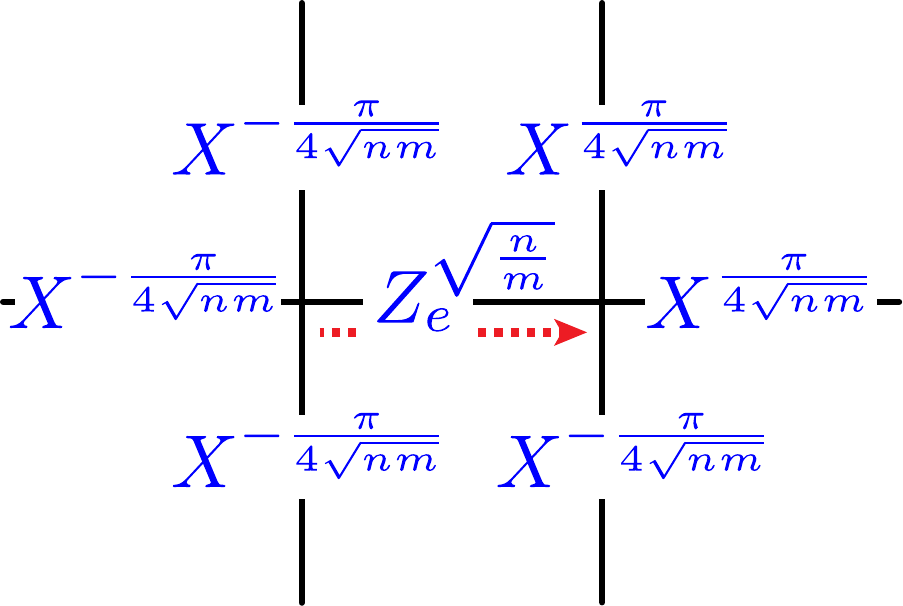}}}\,,\quad \vcenter{\hbox{\includegraphics[scale=0.3]{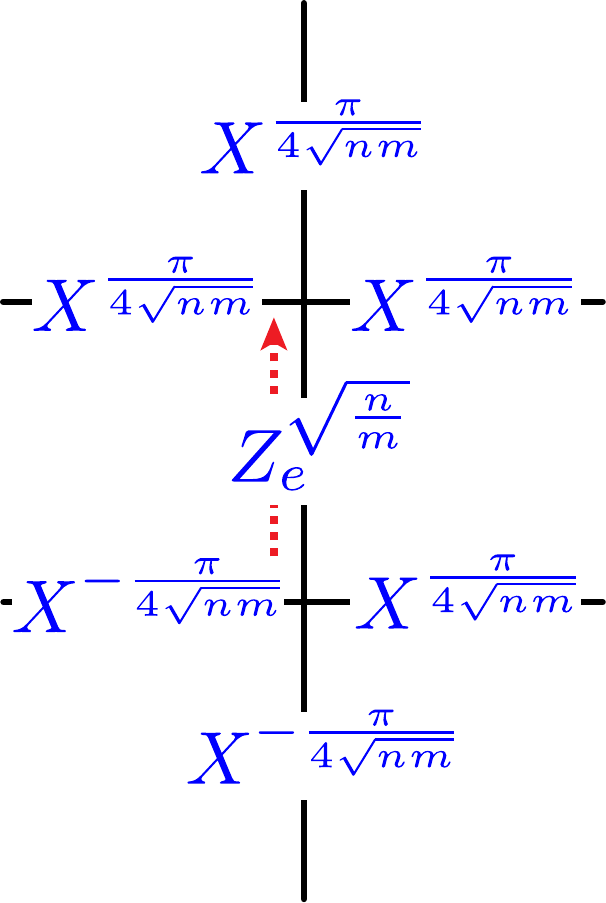}}}.
\end{align}
\end{subequations}
Up to rescaling by a factor of $2$, we recover the operators from Eqs.~\eqref{eq: K24 short strings 2} and \eqref{eq: K24 short strings 4} for $(n,m)=(1,2)$.
In fact, the operators from Eq.~\eqref{eq:twisted-double-condensation-short-strings} generate the full algebra of operators that act within the condensed subspace, i.e., the space in which all operators in $\cB_{n,m}$ act as the identity.
Physically, this corresponds to the limit in which a modified Hamiltonian,
\begin{align}
\begin{split}
    H'_{n,m}(J) =& -\sum_{v\in V} \left(S_v([1,0]) + S_v([0,1])\right)\\
    &- J\sum_{e\in E} \left( C_e\left(2\pi, n \right) +  C_e\left(-2\pi\sqrt{\frac{m}{n}}, \sqrt{nm} \right)\right) + \text{h.c.},
\end{split}
\end{align}
is in the $J\to\infty$ limit.
In that limit, the algebra of operators factorizes into a subalgebra associated to the $\U_{2n}$ factor and another associated to the $\U_{-2m}$ factor. These subalgebras are mutual commutants, analogous to the invertible subalgebras of Ref.~\cite{Haah2023invertible}.

For any finite $J$, there are additional confined excitations. 
For example, non-integer powers of the operators defined in Eq.~\eqref{eq:twisted-double-condensation-short-strings} create confined excitations, as discussed in Section~\ref{sec:example2-4}.
These excitations are confined in the sense that acting on a ground state with a product of the operators along a path results in an energy cost that is extensive in the path length, due their commutation relations with the condensation terms along the path.
In this way, we can create eigenstates of $H_{n,m}$ with arbitrarily small energy, showing that the spectrum is indeed gapless.
In contrast to the single-condensed model in Section~\ref{sec:single-condensation-twisted}, in the present model all topological excitations are gapped and the associated logical ground space is topologically protected, implying a code distance that is extensive in the linear size of the system.

We expect that similar arguments to those presented in Section~\ref{sec: K24 gapping bulk} can be used to show that adding a quadratic perturbation to $H_{n,m}$ opens a gap above the ground space, without changing the ground space properties.

\subsection{Non-chiral even Chern-Simons theories}\label{sec:general_double_condensation}

Lastly, we consider the condensation of a bosonic subgroup defined by a pair of nonzero integers $n_1,n_2\in \bZ\backslash\{0\}$ and an arbitrary integer $n'\in\bZ$ 
\begin{align}\label{eq:general_twisted_cond_bosons}
    \cB = \left\langle (2\pi, n_1), \left( \frac{2\pi n_2}{c_2}, c_2 \right)\right\rangle,
\end{align}
where we have introduced $c_2 = n'+ \sqrt{n'^2-n_1n_2}$.
In order for this to label a valid $\bR$-charge, the following must hold: $n'^2>n_1n_2$.
Additionally, the integers $n_1,n_2,n'$ should be chosen such that $\cB$ is a lattice of dimension 2.\footnote{There are choices of $n_2,n'$ such that $(2\pi n_2/c_2, c_2) \in \langle (2\pi, n_1)\rangle$ or vice versa. In that case, $\cB$ would only need a single generator and hence correspond to a single condensation.}
After condensation the set of deconfined excitations correspond to solutions of the equations
\begin{align}
    b((2\pi, n_1), (x,y)) =2\pi k_1,\quad 
    b\left(\left( \frac{2\pi n_2}{c_2}, c_2 \right), \left(x,y \right) \right) =2\pi k_2, \qq{with} k_1, k_2\in \ZZ.
\end{align}    
Let us denote the  solutions (i.e., the deconfined excitations) by
\begin{align}
    \cA_{\cB} = \left\{ [k_1,k_2]\;|\; k_1,k_2 \in \bZ \right\} \simeq \bZ\times \bZ,
\end{align}
where we have introduced the shorthand notation
\begin{align}\label{eq:evenCS_example_label}
    [k_1,k_2] := \left( \frac{\pi (k_2-k_1 n_2/c_2)}{n'-n_1n_2/c_2}, \frac{k_1(2n'-n_1n_2/c_2) - k_2n_1}{2(n' - n_1n_2/c_2)}\right).
\end{align}
By expressing the generators of $\cB$ in terms of $[k_1,k_2]$ we find that the fusion group describing the inequivalent topological excitations in the condensate is given by the quotient
\begin{align}
    \faktor{\bZ^{\times 2}}{\langle [2n_1,2n'], [2n',2n_2]\rangle } \simeq \bZ_{a}\times \bZ_b,
\end{align}
where $a,b\in 2\bZ$ are positive even integers such that $ab = 4(n'^2 - n_1n_2)$. The explicit values of $a$ and $b$ depend on number-theoretic properties of $n_1,n_2,n'$.
This structure of the topological excitations agrees with the anyon theory of a $\U\times \U$ Chern-Simons theory with $K$-matrix
\begin{align}
    K_{n_1,n_2,n'} = 2\mqty(n_1 & n'\\ n' & n_2).
\end{align}
Because ${n'}^2>n_1n_2$, $K_{n_1,n_2,n'}$ corresponds to an arbitrary non-chiral Chern-Simons theory with an even $K$-matrix.
The braiding of excitations in the condensed phase, written in the $[k_1,k_2]$ basis, is
\begin{align}
\begin{split}
    b([k_1,k_2], [k_1',k'_2]) =\pi \frac{(k_1'k_2+k_1k_2')n'-k_2k_2'n_1-k_1k_1'n_2}{{n'}^2-n_1n_2} =2\pi(k_1,k_2)K^{-1}_{n_1,n_2,n'}\begin{pmatrix}
        k_1'\\
        k_2'
    \end{pmatrix}.
\end{split}
\end{align}
Note that we express the braiding in the $[k_1,k_2]$ basis, obtained from inverting $K$, and not in the basis of the anyons of the model.
The ``anyon basis'' depends on number-theoretic relations between $n_1,n_2$ and $n'$.

\subsubsection{Lattice implementation}
Consider the group of hopping terms associated to $\cB$,
\begin{align}
    \cB_{n_1,n_2,n'} = \left\langle \left\{ C_e(2\pi, n_1), C_e\left( \frac{2\pi n_2}{c_2}, c_2 \right)\right\}_{e \in E} \right\rangle,
\end{align}
where $c_2$ is defined as in Eq.\,\eqref{eq:general_twisted_cond_bosons}.
It is straightforward to verify that the above is a commuting subgroup of displacement operators.
We construct the topological stabilizer group by taking the centralizer over $\cS_{\mathbb{R}}$,
\begin{align}
    C_{\cS_{\bR}}(\cB_{n_1,n_2,n'})  = \left\langle \left\{S_v([1,0]), S_v([0,1])\right\}_{v \in V}\right\rangle,
\end{align}
where we have used the shorthand notation for the labels of the deconfined excitations $[k_1,k_2]$ from Eq.~\eqref{eq:evenCS_example_label}.
Taken together, we can model the condensed phase with the topological code defined by the stabilizer group
\begin{align}
    \cS_{n_1,n_2,n'} =& \left\langle C_{\cS_{\bR}}(\cB_{n_1,n_2,n'}), \cB_{n_1,n_2,n'} \right\rangle\\
            =& \left\langle \left\{S_v([1,0]), S_v([0,1])\right\}_{v \in V}, 
            \left\{ C_e\left(2\pi, n_1 \right), C_e\left( \frac{2\pi n_2}{c_2}, c_2 \right)\right\}_{e \in E} \right\rangle.
\end{align}
Note that the set of generators above is over-complete.
In particular, there exist products of $C_e$ terms on edges along (cellular) coboundaries that multiply to certain $S_v$ operators.
A minimal set of generators is obtained by only including $S_v([k_1,k_2])$ terms corresponding to a minimal set of representatives of the cosets of inequivalent topological excitations $\cA_{\cB}/\cB$.

From the commutation relations between the hopping terms $C_e$ and the logical operators of $\cS_{\bR}$ it follows that the group of logical operators is associated to the group of deconfined excitations after condensation.
For a detailed description of the commutation relations of $C_e$ operators with the logical string operators we refer to Appendix~\ref{app:twisted_hoppingterms}.

Following the examples above, we define a Hamiltonian
\begin{align}
    H_{n_1,n_2,n'} = -\sum_{v\in V} \left(S_v([1,0]) + S_v([0,1])\right)
    -\sum_{e \in E} \left( C_e\left(2\pi, n_1 \right) + C_e\left( \frac{2\pi n_2}{c_2}, c_2 \right)\right) + \text{h.c.}
\end{align}
Since the Hamiltonian terms generate the topological stabilizer group $\cS_{n_1,n_2,n'}$, the ground space coincides with the code space. 
The ground state degeneracy on a torus coincides with the number of anyons, $4(n'^2-n_1n_2)$.
All deconfined excitations are gapped and are created by operators that violate $S_v$ terms in the Hamiltonian and commute with the $C_e$ terms along their length.
Operators that violate the $C_e$ terms create additional confined, gapless, excitations.
Again, following similar arguments to those presented in Section~\ref{sec: K24 gapping bulk}, we expect that a quadratic perturbation to $H_{n_1,n_2,n'}$ is sufficient to open a constant energy gap.

\section{Conclusion and discussion}
\label{sec:Discussion}

In this work, we have used boson condensation to construct a rich set of topological stabilizer codes on continuous variables, starting from a lattice model for a two-dimensional $\bR$ gauge theory.
By condensing pure fluxes and charges in the $\bR$ gauge theory, we recover existing families of topological codes on CVs, such as homological rotor codes~\cite{vuillot2023homological} and codes obtained by concatenating qudit topological codes with (single-CV) GKP codes~\cite{Vuillot2019toricGKP}.
By considering a suitable set of bosonic hopping operators on the lattice, we construct code families that go beyond these existing examples in the literature.
In particular, we find families of topological CV stabilizer codes that support anyon theories that do not admit gapped boundaries. 
This implies that at least one of the following is true:
    \begin{enumerate}
    \item There exist topological stabilizer codes on CVs, including our $K_{2,-4}$ example, that cannot be realized via concatenating a qudit stabilizer code with a local qudit-into-CV encoding. 
    \item There exist topological stabilizer codes on qudits that do not admit gapped boundary conditions. 
    \end{enumerate}
We note that there have been recent developments in Ref.~\cite{ruba2025wittgroups} toward rigorously prove that topological stabilizer codes on discrete variables necessarily admit gapped boundaries, supporting the conjecture that our CV codes are beyond concatenation.

For the subset of our codes characterized by finite anyon theories, we fully expect that they exhibit a threshold to local displacement noise and faulty repeated measurements. 
The reason for this expectation is that the deconfined excitations of the underlying Hamiltonian are discrete. This allows us to leverage decoders for topological codes on discrete variable, which are known to exhibit a threshold~\cite{Bravyi_2013, Anwar2014, Watson2015fast}.
The decoding of our CV codes is thus composed of two stages.
In the first stage, the decoder cleans up the confined excitations, i.e., the violations of $C_e$ terms. 
The confined excitations can be cleaned up in a single-shot and performed in between rounds of measuring the $A_v$ terms. This process is analogous to the first stage of the decoder of the toric-GKP code in Ref.~\cite{Vuillot2019toricGKP}.
The remaining syndromes correspond to violations of $A_v$ terms and represent discrete anyons.
Therefore, the second stage is equivalent to decoding a topological code on discrete variables. In general, this can be performed with a clustering decoder.
Despite our codes not admitting an obvious concatenation structure, the two stage decoding is analogous to the proposed decoding strategy for the toric-GKP code and other concatenated codes.
Given the similarity of the decoding strategy, it is reasonable to assume that the threshold of our CV codes would comparable to that of the toric-GKP code under local stochastic displacement noise and faulty measurements.

Our work reveals unexplored possibilities for CV codes and motivates further investigations into the theory of many-body CV systems, more generally. Regarding the classification of CV stabilizer codes, there are a number of open questions raised by our work.
First, is it possible to prove that all qudit stabilizer codes can be described by anyon theories that admit a gapped boundary? A positive answer to this question would be enough to prove that our codes are beyond concatenated qudit codes. 
Second, which anyon theories can be realized by boson condensation in multiple layers of $\bR$ gauge theory? Here, we have focused on topological codes built from a single layer, but one could consider condensing composite excitations between layers.
Third, are there other classes of topological CV codes that are intrinsic to CV systems, i.e., that are beyond concatenating small CV codes with a qudit code?

Since single-mode GKP codes have been demonstrated experimentally in both superconducting circuits and trapped ions~\cite{PhysRevLett.133.050602, Campagne_Ibarcq_2020}, the realization of multi-mode generalizations appears feasible.
In contrast to other multi-mode code families that are believed to be beyond concatenation~\cite{Lin2023Closest, Harrington2001Achievable, Conrad2022}, our code families are defined by local, translation-invariant stabilizer constraints on a two-dimensional lattice.
This structure makes them promising candidates for intrinsically-CV multi-mode codes, and it would be valuable to explore explicit gadgets to implement our codes experimentally.

Furthermore, we have derived local Hamiltonians from the topological codes. Importantly, we have shown that a gap can be opened in these models by adding quadratic perturbations, which naturally arise in physical realizations of the CV systems. 
From the perspective of QEC codes, these perturbations can be thought of as imposing a soft cutoff on the ground state wave functions, so that the code states become normalizable. 
It is then an interesting problem to understand the ground state wave functions of the perturbed models as approximate code states with finite energy and squeezing.

Small instances of our topological CV codes could be used to design novel topologically protected superconducting qudits, as in Ref.~\cite{vuillot2023homological}. This process would proceed by designing a superconducting circuit whose Hamiltonian admits the codes states as approximate ground states. This provides topological protection for a qudit if there is a (tunable) energetic gap to the excited error states. As discussed in Section~\ref{sec: K24 gapping bulk}, small quadratic perturbations would be sufficient to create such a gap. While small instances of the Hamiltonian models might not offer low-enough logical error rates for an entire computation, they have the potential to achieve lower physical error rates and thereby decrease the overhead of an error-correcting code built using them.

Indeed, beyond QEC codes, this work provides controllably-solvable, and physically reasonable models that exhibit the anyons of a $\U_{2n}$ CS theory. Given that these CS theories are chiral or can be stacked to give models with ungappable boundaries, it is interesting to ask: what boundary theories can be constructed for these Hamiltonians? One might expect to be able to explicitly identify the characteristic low-energy edge modes. Further, can the bulk topological order of our models be diagnosed using standard methods -- for example, what is the topological entanglement entropy of the ground states? In connection to lattice gauge theories, in upcoming work~\cite{CSpaper}, we elaborate on the relations between the Hamiltonians in this work and previous constructions of CS theories on a lattice, e.g., Refs.~\cite{ELIEZER1992118,ELIEZER199266,SunCS2015,Jacobson:2023cmr,Jacobson:2024hov}. Moving beyond topological order, it may be interesting to study whether similar techniques can be used to produce stable, gapped models for fractons in two dimensions~\cite{PhysRevB.96.125151,PhysRevB.97.085116,Williamson2018Fractonic,Radzihovsky2019Fractons}. We are looking forward to exploring these questions in future works.

Finally, a key component of our construction is a generalization of the invertible subalgeras introduced in Ref.~\cite{Haah2023invertible}.
For systems of prime-dimensional qudits, in particular, there is already an established relation between invertible subalgebras and the classification of nontrivial quantum cellular automata (QCAs)~\cite{Haah2023invertible, Freedman2020Classification}. However, a fundamental assumption in the conjectured classification of three-dimensional QCAs is that there does not exist a commuting projector Hamiltonian for anyon theories with ungappable boundaries~\cite{Haah2019CQCA,Haah2023QCA, Shirley2022QCA}. Given that our construction yields two-dimensional stabilizer models with anyon theories that do not have gapped boundaries, it is natural to ask if their existence has implications for QCAs on CV systems.
In particular, do nontrivial QCAs associated to non-chiral anyon theories with ungappable boundaries become trivial as QCAs on CVs?

\vspace{0.2in}
\noindent{\it Acknowledgements -- } 
We grateful to Victor Albert for inspiring discussions about continuous variable stabilizer codes during the early stages of this work. 
TDE and MC thank Jing-Yuan Chen for valuable comments on connections to the lattice gauge theory literature. MC thanks Michael Levin for enlightening conversations about CV Hamiltonians.
DJW thanks Guanyu Zhu and Vladimir Calvera for useful discussions about related ideas.
JM wants to thank J. Conrad for insights into the theory of GKP codes.
We thank Joseph Iosue and Bowen Yang for discussing their unpublished results. 
Part of this work was done during the TOPO 2023 workshop.
JM is supported by the DFG (CRC 183).
Research at Perimeter Institute is supported in part by the Government of Canada through the Department of Innovation, Science and Economic Development and by the Province of Ontario through the Ministry of Colleges and Universities. 
Part of this work was done while DJW was visiting the Simons Institute for the Theory of Computing. Part of this work was done while DJW was based at IBM Almaden Research Center, San Jose, CA 95120, USA. 
DJW was supported in part by the Australian Research Council Discovery Early Career Research Award (DE220100625). MC is supported by NSF grant DMR-2424315.

\begin{appendix}

\section{The algebra of open composite string operators}\label{app:twisted_hoppingterms}

In this appendix, we characterize the algebra generated by short string operators $C_e(\varphi, c)$ used to construct the lattice models in Section~\ref{sec:Examples}.

\subsection{A choice of composite short string operators}

In Eq.~\eqref{eq:def_Wehoppingterms}, we defined the $\bR\times\bR$-labeled operators $\{C_e(\varphi, c)\; {|\; \varphi, c \in \bR\}}$ for every edge $e$ of a square lattice, which we rewrite here for convenience:
\begin{align}
    C_e(\varphi, c) = \vcenter{\hbox{\includegraphics[scale=0.3]{generalCeh.pdf}}}\,,\quad \vcenter{\hbox{\includegraphics[scale=0.3]{generalCev.pdf}}}.
\end{align}
For a fixed edge, $C_e$ is a map $C_e: \bR\times \bR \to \bL(\cH)$, where $\bL(\cH)$ denotes the linear operators on $\cH$. Furthermore, $C_e$ defines a unitary representation of $\bR^{\times 2}$ since
\begin{align}\label{eq:W_onsite_mult}
    C_e(\varphi, c)C_e(\varphi', c') = C_e(\varphi+\varphi', c+c') ,
\end{align}
holds for all $(\varphi, c),(\varphi', c')\in\bR^{\times 2}$.
Since it is a subalgebra of $\bL(\cH)$ it is an associative, unital, $\ast$-algebra. 
Here the $\ast$ operation is given by $\dagger$.
Moreover, the fact that the above operators form a linear representation of $\bR^{\times 2}$ under addition implies that the algebra is commutative.

We want to describe arbitrary linear combinations of (subsets of) the operators $C_e$.
For that, we define $\mathcal{W}_e = \spn_\bC( C_e(\varphi, c)\;|\; \varphi, c \in \bR)$ for each edge.
We interpret $\cW_e$ as a commutative, graded algebra\footnote{We use the direct sum symbol loosely here, as the index set is not discrete and not compact. 
For our purposes, however, it makes sense as elements in the algebra that correspond to bounded operators in $\bL(\cH)$ can be expressed as integrals over displacement operators with coefficients that obey a physicality condition, i.e. they fall off sufficiently quickly for large $(\varphi, a)$. Note that we have not defined a norm on the above algebra in this work.}
\begin{align}
    \cW_e = \bigoplus_{(\varphi,c)\in\bR} V_{(\varphi, c)}^e,
\end{align}
where $V_{(\varphi, c)}^e = \spn_\bC(W_e(\varphi, c))\simeq \bC$.

The algebra of all string operators is generated by products of operators in $\cW_e$ for all edges and can be thought as a tensor product space
\begin{align}\label{app:eq:decompositionWalgebra}
    \cW = \bigotimes_{e\in E} \cW_e = \bigotimes_{e\in E} \bigoplus_{\varphi,c\in\bR} V_{(\varphi, c)}^e,
\end{align}
where the tensor product corresponds to the multiplication of operators in $\bL(\cH)$.
Note that this tensor product is not commutative since the underlying operators don't commute in general.

\subsubsection{String operators form an operator basis}
We now confirm that any element in $\bL(\cH)$ can be represented by an element in $\cW$ and vice versa, any element in $\cW$ is a linear operator on $\cH$ by construction.
This implies that they are the same algebra, $\bL(\cH) =\cW$.
To see this, we show that $\cW$ contains a complete basis of $\bL(\cH)$, namely the single-edge displacement operators.
Single-CV displacement operators can be written as a product of $X^\varphi$ and $Z^c$ operators defined in Eq.~\eqref{eq:displacements-def}, for $\varphi, c\in \bR$ and a complex phase.
We now identify these operators, on an arbitrary edge $e$, with elements in $\cW$.
The $X$-like displacements are easily generated.
For each edge,
\begin{align}
    C_e(\varphi, 0) = X_e^\varphi.
\end{align}
To identify the $Z$-like displacements as elements in $\cW$, we have to combine the generators $C_e(\varphi, c)$ in a nontrivial way since $C_e(0,c)$ acts nontrivially on 6 CVs.
First, we note that taking an appropriate product of $C_e(\pm\varphi, \pm c)$, on edges surrounding a vertex $v$, we obtain trivial-loop operators $S_v(\varphi, c)$ defined in Eq.~\eqref{eq:composite-R-stabilizer}.
These operators, in fact, define a commutative subalgebra of $\cW$ that is described in more detail in Appendix~\ref{sec:loop-algebra}.
Setting $\varphi = 0$, the resulting operator only acts on CVs around a single plaquette, specifically, the plaquette to the north-west of vertex $v$.
We can hence multiply $C_e(0,c)$ by $S_v(0,c/2)$ for a neighboring vertex $v$ and obtain,
\begin{align}
    C_e(0,c) S_v(0,c/2)= Z_{e'}(c) .
\end{align}
Since we can perform this mapping for any edge $e'$, we find that all displacement operators are contained in $\cW$ and with that $\cW=\bL(\cH)$.

Having established that $\cW$ is the same algebra as the full algebra generated by single-CV displacements, we continue our discussion of $\cW$.
Since the objects in $\cW_e$ and $\cW_{e'}$ can have nontrivial overlap (as operators on $\bL(\cH)$), they might not commute. 
This defines a braiding in terms of the basis,
\begin{align}\label{eq:psi_def}
C_e(\varphi, c) C_{e'}(\varphi', c') = \psi_{e,e'}((\varphi, c), (\varphi', c')) C_{e'}(\varphi', c') C_e{(\varphi, c)},
\end{align}
where $\psi_{e,e'}$ assigns a complex phase to each tuple in $\bR^{\times 2}\times\bR^{\times 2}$.
Since $C_e$ forms a unitary representation of $\bR^{\times 2}$ on each edge individually, it follows that $\forall g,h\in\bR^{\times 2}$
\begin{align}
    \psi_{e,e'}(g,h) = \psi_{e',e}(-h,-g)^\ast \qq{and}
    \psi_{e,e'} (g,h)\psi_{e,e'} (g',h) = \psi_{e,e'}(g+g',h),
\end{align}
where $\pm$ denotes the addition (subtraction) in $\bR^{\times 2}$ and $\ast$ denotes complex conjugation.
From these two properties we can infer that the second line holds analogously for the second argument.
As such, $\psi$ is given by a bilinear form on $\bR^{\times 2}$.\footnote{In fact, it is the symplectic form on the vector space spanned by the displacement operators modulo phases, in the basis defined by $W_e$.}

In the following, we calculate $\psi$ explicitly and find that it is connected to the exchange statistics of the charges and fluxes in the $\bR$ gauge theory.
First, we observe that
\begin{align}
    \psi_{e,e} \equiv 1\qcomma \psi_{e,e'}((0,0), (\varphi, c)) = 1, \,\,\forall (\varphi, c) \qq{and} \psi_{e,e'} = (\psi_{e',e})^\ast.
\end{align}
Moreover, we find that $\psi_{e,e'} \equiv 1$ for all pairs of edges except these cases
\begin{align}
    \psi_{e,e'} ((\varphi,c), (\varphi',c')) = \begin{cases}
        e^{i(\varphi c' + \varphi' c)/2} & \raisebox{-0.4\height}{\includegraphics[height=24pt]{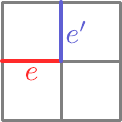}}, \raisebox{-0.4\height}{\includegraphics[height=24pt]{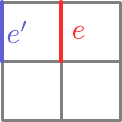}}\\[12pt]
        e^{-i(\varphi c' + \varphi' c)/2} & \raisebox{-0.4\height}{\includegraphics[height=24pt]{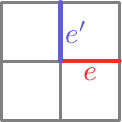}}, \raisebox{-0.4\height}{\includegraphics[height=24pt]{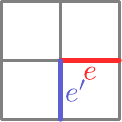}}, \raisebox{-0.4\height}{\includegraphics[height=24pt]{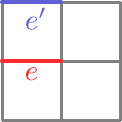}}, \raisebox{-0.4\height}{\includegraphics[height=24pt]{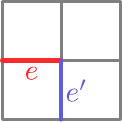}}
    \end{cases}.
\end{align}
Note that $\psi_{e,e'}$ is uniquely defined by a bilinear form $b((\varphi, c), (\varphi',c')) = (\varphi c' + \varphi' c)/2$.\footnote{Bilinearity can be checked straight forwardly,
\begin{align*}
    b((\varphi,c)+ (\varphi',c'), (\vartheta, b)) =& \frac{1}{2}((\varphi + \varphi')b + \vartheta(c+c'))
    = \frac{1}{2}(\varphi b + \vartheta c) + \frac{1}{2}(\varphi' b + \vartheta c')\\
    =& b((\varphi, c), (\vartheta, b)) + b((\varphi',c'), (\vartheta, b)),
\end{align*}
which holds similarly in the second argument.}
From now on, we refer to $\psi$ informally as the braiding of $\cW$.
We find that the braiding fully characterizes the algebra.
For example, a T-junction calculation~\cite{Levin2003Fermions, Ellison2022subsystem} shows that the exchange statistics of the anyons that are created and moved by elements in $\cW$ is given by
\begin{eqs} \label{eq:top_spin_shortstrings}
    \theta\Big((\varphi, c)\Big) =& \psi_{e,e'}((\varphi, c), (-\varphi, -c)) \psi_{e,e''}((\varphi, c), (-\varphi, -c)) \psi_{e',e''}((-\varphi, -c), (-\varphi, -c))
    = e^{i\varphi c}.
\end{eqs}

\subsection{Dictionary to cellular homology}

The operators $C_e(\varphi, c)$ are naturally associated to 1-cochains with $\bR^{\times 2}$ coefficients on a cellulation $\Delta$ (a cell complex with 0-chains $V$, 1-chains $E$ and 2-chains $P$). Note that throughout this paper, we work on a square lattice.

We first construct a chain complex of real vector spaces from $\Delta$ and $\bR^{\times 2}$.
Namely, we define $C(\Delta, \bR^{\times 2})$ to be the chain complex with the $i$-chains being vectors in
\begin{align}
    C_i = \spn_{\bR\times \bR}(\Delta_i).
\end{align}
Since $\Delta$ is equipped with an orientation, we impose that orientation reversal in the complex corresponds to inverting in $\bR^{\times 2}$, i.e.
\begin{align}
    \alpha c^{-1} = -\alpha c\qcomma \alpha\in \bR^{\times 2}, c\in \Delta_i.
\end{align}
The simplicial complex is equipped with a notion of boundary obtained by mapping each $i$-cell to a formal sum over $(i-1)$-cells that have a certain overlap with the original $i$-cell, taking orientations into account.
In our case, we can interpret this boundary as summing over the ${(i-1)}$-cells in the geometric boundary of a given $i$-cell, taking the orientation inherited from the $i$-cell into account.
For example, the boundary of a 2-cell can be interpreted as summing over the surrounding edges in counterclockwise orientation.
For a more thorough treatment, we refer to Ref.~\cite{algebraictopology}.
Explicitly, we obtain a chain complex by defining the boundary operator $\partial_i: C_i\to C_{i-1},\; v_i \mapsto \sum_{s\in \partial v_i} s$ where $v_i\in \Delta_i$, and extending it linearly, i.e.
\begin{align}
    \partial_i (\alpha v + \beta w) = \alpha \partial_i(v) + \beta \partial_i(w)\qcomma \forall \alpha,\beta\in \bR^{\times 2},\; v,w\in \Delta_i .
\end{align}
Below, we drop the subscript $i$ if it is clear from context what chains $\partial$ it acts on.
From this definition (with orientation dependent $\pm$ signs included) it is clear that $\partial_{i-1}\circ\partial_i = 0$.
Elements $\ker(\partial_i)$ are called \textit{$i$-cycles} and the group of all i-cycles is denoted by $Z_i(\Delta,\bR^{\times 2})$.
Elements in $\Im(\partial_{i+1})$ are \textit{$i$-boundaries} and the group of all i-boundaries is denoted by $B_i(\Delta,\bR^{\times 2})$.
The homology groups of the complex are 
\begin{align}
    H_i(\Delta,\bR^{\times 2}) = \faktor{Z_i(\Delta,\bR^{\times 2})}{B_i(\Delta,\bR^{\times 2})}.
\end{align}

In the construction above we can take the dual complex to obtain cochains, cocycles, coboundaries, and cohomology groups.
Since we only consider finite cellulations, each $C_i$ is a finite-dimensional vector space and $\partial$ is represented by a matrix.
The coboundary operator $\delta$ in this case is $\partial^T$ and represents homology on the Poincaré dual cellulation.
Since we have a finite-length complex, derived from a finite-dimensional manifold, i.e.~$C_j = 0$ for $j\neq 0,1,2$ there is a natural identification
$C_0\simeq C^2$, $C_1\simeq C^1$ and $C_2\simeq C^0$ as vector spaces.
In the following it is useful to consider cochains.

From the perspective introduced above one might be inclined to interpret the collection of $C_e$ as defining a map from 1-cochains $C^1(\Delta,\bR^{\times 2})$ to $\bL(\cH)$.
In order for this map to be well-defined, however, we would need to specify an order of multiplication since the operators in the image of $C_e$ and $C_{e'}$ do not necessarily commute in $\bL(\cH)$, but the addition in $C^1(\Delta,\bR^{\times 2})$ is commutative.
This can be solved by, for example, defining an arbitrary global ordering on $\Delta^1$.
Since this is not necessary to capture the essential phenomena, we take a different approach.
We do not aim to map onto individual operators in $\bL(\cH)$ but into subspaces spanned by operators in the set of all linear subspaces of $\bL(\cH)$.
Here, we consider subspaces spanned by a group of displacement operators $D$.
Since they commute up to a phase this is equivalent to mapping into the group modulo phases $\hat{D} = D/\sim$, where $d_1\sim d_2$ if $\exists \varphi\in [0,2\pi): d_1 = e^{i\varphi} d_2$.
Note that $\hat{D}$ is a real, symplectic vector space isomorphic to $\bR^{2n}$, where $n$ is the total number of modes, where the symplectic form $\lambda: \hat{D}\to \bR$ is defined by the commutation relation among displacements, i.e. $d_1d_2 = e^{i\lambda(d_1,d_2)}d_2d_1$.

Let $\mathcal{D}$ be the space of all subspaces in $\bL(\cH)$ that are spanned by displacement operators.
We define the map $\Xi: C^1(\Delta, \bR^{\times 2}) \to \mathcal{D}$,
\begin{align}
    c = \sum_e \alpha_e e \mapsto \spn_\bC \left(\prod_{e}C_e(\alpha_e) \right). 
\end{align}
Note that $\Xi$ is a group homomorphism, where the group operation on $\mathcal{D}$ is the tensor product defined below Eq.~\eqref{app:eq:decompositionWalgebra} via multiplication of operators.
It follows directly that the order of multiplication does not matter for the mapping, as expected.

We can extend $\psi$ from Eq.~\eqref{eq:psi_def} linearly onto to equip elements in $\mathcal{D}$ with an algebra structure.
In the following, we describe subalgebras in $\bL(\cH)$ as images of subgroups of $C^1(\Delta,\bR^{\times 2})$ under $\Xi$.

\subsection{Algebra of coboundaries -- $\bR$ gauge theory}\label{sec:loop-algebra}

Consider the subgroup of all 1-coboundaries $B^1(\Delta,\bR^{\times 2}) = \Im(\delta_0)$.
We define a space of operators
\begin{align}\label{eq:coboundary_algebra}
    \cS = \Im(\Xi|_{B^1(\Delta,\bR^{\times 2})}).
\end{align}
For example, for a vertex $v\in\Delta_0$, there is a 1D-subgroup in $B^1(\Delta,\bR^{\times 2})$ of the form $\alpha_v v$ for $\alpha_v\in\bR^{\times 2}$.
The image of this subgroup under $\Xi$ is the one-dimensional space $\cS_v(\varphi,c)$  spanned by $S_v(\varphi, c)$, defined in Eq.~\eqref{eq:composite-R-stabilizer}.
Since we included a subgroup of $C^1(\Delta,\bR^{\times 2})$ we find that $\cS$ is a subalgebra of $\bL(\cH)$, i.e. closed under multiplication.
It is spanned by all possible products of operators $S_v(\varphi, c)$ for all values of $\varphi, c$ and vertices $v$.
We find that two such operators commute
\begin{align}
    \comm{S_v(\varphi, c)}{S_{v'}(\varphi', c')}_G = \mathds{1}\qcomma\forall v,v'\in V,\; \varphi,\varphi',c,c'\in \bR,
\end{align}
showing that $\cS$ is commutative. As a shorthand we write $\psi|_\cS = 1_\cS$.
Hence, we refer to $\cS$ as a stabilizer algebra and the unitaries $\{S_v(\varphi, c)\}_{(\varphi, c), v}$ as stabilizers, as they are the stabilizers of the $\bR$ gauge theory model in Section~\ref{sec: R gauge theory model}.

More concretely, any element in $\cS$ can be written as
\begin{align}\label{eq:stabilizer_cont_expansion}
    S(\vb{c}) = \int_{\bR^{\times 2}}\dd{(\varphi, c)} \prod_{v\in V} \vb{c}_v(\varphi, c) S_v(\varphi, c),
\end{align}
where the vector of functions $\vb{c}: \bR\times \bR \to \bC^V$ encodes the coefficients\footnote{We do not make any assumptions on $\vb{c}$ here. This allows unbounded operators and unphysical density matrices in $\cS$. However, we can still make formal calculations in that algebra given the structure as described above.} in front of the generators $S_v(\varphi, c)$.
$\vb{c}_v(\varphi, c)$ denotes the $v$th entry of the image of the function $\vb{c}$, evaluated at $(\varphi, c)$.

We now consider the commutant of $\cS$, over $\cW$,
\begin{align}
    \cC_\cW(\cS) = \{w\in\cW\;|\; ws=sw\;\forall s \in \cS\}.
\end{align}
We find that any open strings, i.e., operators in the image of $\Xi|_{C^1(\Delta,\bR^{\times 2})- Z^1(\Delta,\bR^{\times 2})}$ do not commute with $\cS$.
At the same time, the algebra associated to cocylces commute with $\cS$.
Trivial loops are elements of $\cS$, and hence commute with $\cS$ as shown above.  Products along nontrivial cycles also commute with $\cS$, as we now show explicitly. 
Let $\gamma\in Z^1(\Delta,\bR^{\times 2})$ be a 1-cocycle and let $W_\gamma$ denote an operator in the one-dimensional subspace $\Xi(\gamma)$.
We find
\begin{align}
    \comm{S_v(\varphi,c)}{W_\gamma}_G = \mathds{1}\qcomma \forall v\in V, \gamma\in Z^1(\Delta, \bR^{\times 2})\qq{and} \varphi,c\in \bR.
\end{align}
This can be checked by picking a basis of $Z^{1}(\Delta,\bR^{\times 2})$ and explicitly computing the commutation relation.
Note that any $w\in\cW$ can be written as a sum of terms that only involve products over cocycles (cochains with trivial coboundary) and a sum of terms that only involve products over cochains with nontrivial coboundary. Moreover, the subspace spanned by products along cocycles is closed under multiplication.
Hence, we have found all elements of $\cW$ in the commutant and we can write
\begin{align}
    \cC_\cW(\cS) = \Im(\Xi|_{Z^{1}(\Delta,\bR^{\times 2})}).
\end{align}
We can separate the cohomology classes and find that within each cohomology class, the spaces $\Xi(\gamma)$ and $\Xi(\gamma')$ for two cohomological $\gamma,\gamma'$ differ by multiplication of an element in $\cS$.
Since $\cS$ is central in $\cC_\cW(\cS)$ the commutation of operators in $\Xi(\gamma)$ and $\Xi(\gamma')$ within $\cC_\cW(\cS)$ only depend on the cohomology class $[\gamma]=[\gamma']$.
More concretely, take two one-dimensional subspaces $\Xi(\gamma)$ and $\Xi(\eta)$ assigned to two cocycles $\gamma,\eta \in Z^1(\Delta,\bR^{\times 2})$.
Let $O_\gamma\in\Xi(\gamma)$ and $O_\eta\in \Xi(\eta)$. We can evaluate $\psi$ on these $\gamma$ and $\eta$ to obtain a complex phase $\psi(\gamma,\eta)$ such that
\begin{align}
        O_\gamma O_\eta = \psi(\gamma,\eta) O_\eta O_\gamma.
\end{align}
What we find is that $\psi|_{Z^1(\Delta,\bR^{\times 2})^{\times 2}}$ only depends on the cohomology class of the arguments.

For example, for a torus $H^1(\Delta_{\mathbb{T}^2},\bR^{\times 2}) = \bR^{\times 2}\times \bR^{\times 2}$. We can identify a natural basis for this group in which the cohomology class is defined by two coefficients $\alpha,\beta\in\bR^{\times 2}$, each of which is assigned to one of the two generating cocycle classes.\footnote{We can write each cocycle class as $\alpha [\gamma] + \beta[\gamma']$ where $[\gamma]\neq[\gamma']$ and $\alpha,\beta\in\bR^{2}$.}
In this basis, $\psi((\alpha,\beta),(\alpha',\beta')) = e^{i(\alpha_1\beta'_2 + \alpha_2\beta'_1 + \alpha'_1\beta_2 + \alpha'_2\beta_1)} = e^{i(\alpha_1\beta'_2 + \alpha_2\beta'_1)} e^{i(\alpha'_1\beta_2 + \alpha'_2\beta_1)}$.
The factorization indicates that the logical algebra factorizes over two isomorphic subalgebras that commute with each other.
In fact, each of these subfactors admits an $\bR^{\times 2}$-indexed basis $\{D(\varphi,c)\;|\; \varphi,c \in\bR\}$ of operators that fulfill
\begin{align}
    D(\varphi, c)D(\varphi',c') = e^{i(\varphi c' + \varphi' c)} D(\varphi',c') D(\varphi, c).
\end{align}
which is the commutation relation of displacement operators. In that sense we find that on a torus, the nontrivial part of the commutant is isomorphic to $\bL(L^2(\bR))^{\otimes 2}$.

The construction above is a reoccurring theme in this work. We first find a commuting subalgebra of $\cW$, calculate its commutant and identify the equivalence classes up to multiplication of elements in its center (which, by construction, includes the commuting subalgebra that we started with).
The commutant can be decomposed into its center which is a potentially infinite product of commuting one-dimensional algebras, and a factor that spans a logical algebra.
In the case encountered in this section we found an infinite-dimensional logical algebra.
Below, we see how different choices of commuting subalgebras can lead to finite-dimensional logical (sub)algebras.

\subsection{Homogeneous subalgebras and conditions on commutativity}\label{sec:homogen_subalg}
We have seen how reducing $\Xi$ onto 1-coboundaries leads to a commuting subalgebra of $\cW = \bL(\cH)$.
We have shown how $\cS$ and its commutant can be obtained by plugging 1-coboundaries and 1-cocycles into $\Xi$.
Through this mapping we found that there is no 1-cochain with a coboundary that commutes with $\cS$ (more precisely, no 1-cochain with a nontrivial coboundary whose image under $\Xi$ commutes with $\cS$). 
In the interpretation in terms of $\bR$ gauge theory, the algebra of operators obtained from cochains with coboundary are {charged operators} that create flux-charge excitations.
Hence, a subalgebra that commutes with (a subalgebra) of these operators, should be thought of as defining a {condensate} in which certain charged operators act trivially.
The goal of this section is to quantify for which subalgebras of that form commute such that we can employ the stabilizer formalism to construct a model for the condensate.
We find that in order to achieve commutativity, we need to restrict to a proper subgroup of $\bR^{\times 2}$.
We first define a general algebra of that form, for any subgroup, and then give conditions on when such an algebra is commutative for different isomorphism classes of subgroups.

\begin{definition}[homogeneous subalgebra]
Let $H\leq \bR^{\times 2}$ be a subgroup.
For each edge, we define an algebra\footnote{Since $H$ is a non-compact group, and can even be continuous, the direct sum symbol should be interpreted loosely. We use it to refer to a formal vector space whose elements can be expressed as group integrals over operators in $V^e_{(\varphi, c)}$, see the expansion in Eq.~\eqref{eq:stabilizer_cont_expansion}.}
\begin{align}
    \cW_e|_H = \bigoplus_{(\varphi,c) \in H} V^e_{(\varphi, c)}.
\end{align}
We call an algebra of the form
\begin{align}
    \cW|_H = \bigotimes_{e\in E} \cW_e|_H,
\end{align}
a \textit{homogeneous subalgebra} of the subgroup $H$.
\end{definition}

We can view this construction in terms of restricting $\Xi$ to a subgroup $C^1(\Delta, H)\leq C^1(\Delta,\bR^{\times 2})$ and identify $\cW|_H = \Im(\Xi|_{C^1(\Delta, H)})\subseteq \cW$. The fact that $H$ forms a group ensures that $\cW|_H$ is indeed an algebra by the group homomorphism property of $\Xi$.

There are finitely many isomorphism classes of subgroups of $\bR^{\times 2}$:
\begin{enumerate}
    \item $H_0\simeq \bZ_0 = \{0\}$
    \item $H_{1,1}\simeq \bR$
    \item $H_{1,2}\simeq \bZ$
    \item $H_{2,1}\simeq \bZ\times \bR$
    \item $H_{2,2}\simeq \bZ\times \bZ$
    \item $H_{3}\simeq \bR^{\times 2}$.
\end{enumerate}
The following cases lead to trivial results $\cW|_{H_0} \simeq \bC$ and $\cW|_{H_3} = \cW$, and hence we only consider the proper subgroups.
In the following, we derive exact conditions for $H$ to define a commuting subalgebra, i.e. for which $\psi|_{\cW_H} = 1|_{\cW_H}$.

\begin{itemize}
    \item $H_{1,1} = \{\alpha (\varphi_0,c_0)\;|\; \alpha\in \bR\}\simeq \bR$: The commutativity condition reads
    \begin{align}
        \alpha\beta \varphi_0 c_0 = 2\pi k\qcomma k\in\bZ, \forall \alpha,\beta\in\bR.
    \end{align}
    Since this condition has to hold for all $\alpha, \beta$, we obtain that either $\varphi_0$ or $c_0$ have to be $0$.
    This mimics the conditions on $H$ that describe a bosonic continuous subgroup of the topological excitation in $\bR$ gauge theory isomorphic to $\bR$, see Eq.~\eqref{eq:R-gauge-spin}.

    \item $H_{1,2} = \{n(\varphi_0,c_0)\;|\; n\in \bZ\}\simeq \bZ$:
    In the case of a finitely generated subgroup with a single generator the commutativity relation reduces to
    \begin{align}
        n m\varphi_0c_0 = 2\pi k\qcomma k\in\bZ, \forall n,m \in\bZ.
    \end{align}
    This is fulfilled either for $\varphi_0$ or $c_0$ equal to $0$ or, assuming $c_0\neq 0$, if and only if
    \begin{align}\label{eq:lattice-singlecondensation-boson}
        \varphi_0 = \frac{2\pi}{c_0} k,
    \end{align}
    for some $k\in\bZ$.
    This coincides with the condition on $H$ that describes a finitely generated subgroup of bosons in $\bR$ gauge theory, see Eq.~\eqref{eq:R-gauge-boson-general}.
    
    \item $H_{2,1} = \{n(\varphi_0,c_0) + \alpha(\varphi_1,c_1)\;|\; n\in \bZ, \alpha\in \bR\}\simeq \bZ\times\bR$:
    If the subgroup admits one finitely generated factor and a continuous one, the commutativity condition reduces to
    \begin{align}
        \alpha\beta \varphi_0c_0 + n m \varphi_1c_1 + \frac{\alpha m + \beta n}{2}(\varphi_1 c_0 + \varphi_0 c_1) = 2\pi k\qcomma k\in\bZ ,
    \end{align}
    for all $\alpha,\beta\in\bR$, and $n,m\in\bZ$.
    Again, since this has to hold for all values of the coefficients, we can deduce that this is equivalent to
    \begin{subequations}
        \begin{align}
            \varphi_0 c_0 = \varphi_0 c_1 + \varphi_1 c_0 =& 0\qq{and}\\
            \varphi_1 c_1 =& 2\pi k'\qcomma k'\in\bZ.
        \end{align}
    \end{subequations}
    Again, we recover the same conditions for $H$ as for a subgroup of bosons in the $\bR$ gauge theory. 
    The first condition enforces that the continuous part of $H$ is a bosonic subgroup that braids trivially with the generator of the discrete part. The second condition enforces that the discrete part is generated by a boson.

    \item $H_{2,2} = \{n(\varphi_0,c_0) + m(\varphi_1,c_1)\;|\; n,m\in \bZ\}\simeq \bZ\times\bZ$:
    So far, we have seen how the condition of exact commutativity of hopping terms on the lattice force the anyons that they hop to be bosonic, in the sense that the associated topological spin is 1, see Eq.~\eqref{eq:R-gauge-spin}.
    We now encounter the first and only case where this is not so. 
    On the lattice, there is an additional constraint on $H$ coming from the commutativity.
    
    Specifically, for subgroups of the form as $H_{2,2}$ the commutativity condition reads
    \begin{align}
        n n' \varphi_0 c_0 + m m' \varphi_1 c_1 + \frac{nm'+n'm}{2}(\varphi_0 c_1 + \varphi_1 c_0) = 2\pi k\qcomma k\in\bZ,
    \end{align}
    for all $n,n'm,m'\in\bZ$.
    Again we deduce that each factor individually has to be generated by a pair $(\varphi_i,c_i)$ that fulfills 
    \begin{align}\label{eq:H22_condition1}
        \varphi_i c_i = 2\pi k_i\qq{with} k_i\in\bZ.
    \end{align}
    This agrees with the pair representing a boson in $\bR$ gauge theory, see Eq.~\eqref{eq:R-gauge-boson-general}.
    The cross terms, however, lead to a slightly different condition
    \begin{align}\label{eq:H22_condition}
        \varphi_0 c_1 + \varphi_1 c_0 = 4\pi \Tilde{k}\qcomma \Tilde{k}\in \bZ.
    \end{align}
    This is equivalent to demanding that the mutual monodromy phase of the excitations associated to the generators of $H_{2,2}$ should be an integer power of $e^{4\pi i}$ instead of $e^{2\pi i}$.
\end{itemize}

The necessary conditions for different choices of $H$ apply only to homogeneous subalgebras where the local algebras are built from operators $C_e(\varphi, c)$.
In other words, the definition of $\Xi$ is basis-dependent.
In general, one can think of changing this basis to a different local algebra that is graded by $\bR\times\bR$. For example, taking single-CV displacements could produce a valid choice.
These, however, leads to homogeneous subalgebras that describe the condensation of pure charges or pure fluxes in the $\bR$ gauge theory, see Section~\ref{sec:condensation} and highlights the importance of picking the right set of condensation terms in the lattice model.

\section{Group structure of the anyons after condensing a composite boson}\label{app:twisted-condensation-group}
In this appendix, we prove that after condensing a discrete subgroup of bosons generated by a nontrivial flux-charge composite in an $\bR$ gauge theory, the set of inequivalent deconfined excitations is of the form described in Eq.~\eqref{eq:1condensation-twisted-anyons}.
We do so by explicitly showing that the labeling of cosets
\begin{align}\label{app:eq:cosets-labels}
    \faktor{\cA_{B}}{\cB} = \left\{\left(q(\pi/n, 1/2) + (\alpha, -\alpha n/2\pi) \right) + \cB\;|\; q\in\bZ_{2n},\alpha\in\bR\right\},
\end{align}
where $\cB = \langle (2\pi, n)\rangle$, defines a group isomorphism
\begin{align}\label{app:eq:isomorphism}
    \bZ_{2n}\times \bR \stackrel{\sim}{\longrightarrow} \faktor{\cA_{B}}{\cB} .
\end{align}
In the following, we introduce the notation $[q,\alpha] = q(\pi/n, 1/2) + (\alpha, -\alpha n/2\pi)$ for $q\in\bZ, \alpha\in \bR$.
Clearly, if we allowed for $q\in\bZ$ in Eq.\,\eqref{app:eq:cosets-labels} the labelling would define a group homomorphism $\bZ\times\bR\to \cA_\cB/\cB$.\footnote{Since the labels $q$ and $\alpha$ enter linearly into the definition of the coset representatives the group operation is preserved.}
In the following we show that it becomes a group isomorphism when  the first factor is restricted to $q\in\bZ_{2n}$.
We do so by explicitly showing that the induced map (see  Eq.\,\eqref{app:eq:isomorphism}) is a bijection.

\paragraph{Surjectivity:} Let $[x,y] + \cB \in \cA_{\cB}/\cB$ be an arbirary coset.
We now show that any such coset can be represented by $[q,\alpha]$ for $q\in\bZ_{2k},\alpha\in\bR$, i.e. we find a $b = [b_1,b_2]\in \cB$ such that 
\begin{align}
    [q,\alpha] = \left[x + b_1, y +b_2 \right],
\end{align}
for a suitable choice of $q,\alpha$.
A straightforward calculation shows that this is achieved by 
\begin{align}
    b_2=0\qcomma \alpha=y\qcomma q - b_1 = x.
\end{align}
Noting that $\cB = \langle [2n,0]\rangle$, we find that the first condition is fulfilled for any $b\in\cB$.
Similarly, $\alpha\in\bR$ allows for the second equation to be fulfilled.
To see that there exist a suitable choice of $q \in \bZ_{2n}$ and $b_1\in 2n\bZ$, we note that $q$ has to be an integer to be in $\cA_\cB$ (see Eq.\,\eqref{eq:def_AB_twistedcondensation}). By a suitable choice of $b_1$, we can always find $q\in\bZ_{2n}$ such that the third condition is fulfilled.

\paragraph{Injectivity:}
Consider two cosets $[x, y] + \cB$ and $[x', y'] + \cB$ with $x,x'\in \bZ_{2n}; y,y'\in\bR$, i.e. the images under the map above for two elements in $\bZ_{2n}\times\bR$.
Since $\cB$ is generated by $[2n,0]$ these two cosets are the same if and only if there exists $\ell\in\bZ:$
\begin{align}
    \left[x, y\right] = \left[x' + 2n\ell, y'\right]
    = \begin{cases}
        x = x' + 2n\ell\\
        y = y'
    \end{cases}.
\end{align}    
Since $x,x'\leq 2n$ it follows that $\ell = 0$ and with that $[x,y] = [x',y']$, proving injectivity.

\end{appendix}

\bibliographystyle{quantum}
\bibliography{bib}

\end{document}